\documentclass[prd,aps,floats,twocolumn]{revtex4}
\usepackage[usenames,dvipsnames]{color}
\usepackage{graphicx, array}
\usepackage{graphics,epsfig}
\usepackage{amssymb}
\usepackage{amsmath}
\usepackage{ulem}
\usepackage{multirow}
\usepackage{subfigure}
\usepackage{ulem}
\usepackage{bm}
\pdfoutput=1
\newcolumntype{C}[1]{>{\centering\arraybackslash}m{#1}}

\def\be{\begin{equation}}
\def\ee{\end{equation}}
\def\bi{\begin{itemize}}
	\def\ei{\end{itemize}}
\def\ben{\begin{enumerate}}
	\def\een{\end{enumerate}}
\def\bt{\begin{tabular}}
	\def\et{\end{tabular}}
\def\bc{\begin{center}}
	\def\ec{\end{center}}

\def\bea{\begin{eqnarray}}
\def\eea{\end{eqnarray}}

\def\ba{\begin{eqnarray}}
\def\ea{\end{eqnarray}}

\def\mnras{Mon. Not. R. Astron. Soc.}
\def\jcap{J. Cosmol. Astropart. Phys.}
\def\aj{Astronomical Journal}
\def\apj{Ap. J.}
\def\apjs{Ap. J. S.}
\def\apjl{Ap. J. Lett.}
\def\doff{d} 

\begin{document}

\input{epsf}
\title{Cluster mislocation in kinematic Sunyaev-Zel'dovich (kSZ) effect extraction }
\author {Victoria Calafut$^{1}$, Rachel Bean$^{1}$ and Byeonghee Yu$^{1,2}$}
\affiliation {$^{1}$Department of Astronomy, Cornell University, Ithaca, NY 14853, USA, \\ $^{2}$Department of Physics,
University of California, Berkeley, CA 94720, USA.} 
\label{firstpage}

\begin{abstract}
We investigate the impact of a variety of analysis assumptions that influence cluster identification and location on the kSZ pairwise momentum signal and covariance estimation. Photometric and spectroscopic galaxy tracers from SDSS, WISE, and DECaLs, spanning redshifts $0.05<z<0.7$, are considered in combination with CMB data from Planck and WMAP.  With two complementary techniques, analytic offset modeling and direct comparisons of redMaPPer brightest and central catalog samples, we find that miscentering uncertainties average to $0.4-0.7\sigma$ for the Planck  kSZ statistical error budget obtained with a jackknife (JK) estimator. We also find that JK covariance estimates are significantly more conservative than those obtained by CMB rotation methods. Using redMaPPer data, we concurrently compare the impact of photometric redshift errors and miscentering. At separations $<\sim 50$ Mpc, where the kSZ signal is largest, miscentering uncertainties can be comparable to JK errors, while  photometric redshifts are lower but still significant. 

For the next generation of CMB and LSS surveys the statistical and photometric errors will shrink markedly. Our results demonstrate that uncertainties introduced through using galaxy proxies for cluster locations will need to be fully incorporated, and actively mitigated, for the kSZ  to reach its full potential as a cosmological constraining tool for dark energy and neutrino physics.
\end{abstract}

\maketitle
\section{Introduction}
\label{sec:intro}

The last decade has seen large scale primordial Cosmic Microwave Background (CMB) temperature and polarization anisotropies measured down to cosmic variance levels, with the Wilkinson Microwave Anisotropy Probe (WMAP) \cite{2013ApJS..208...19H}, and Planck \cite{2016A&A...594A...1P} satellites. Concurrently, the measurement of anisotropies at arcminute scales, for example with Planck \cite{Ade:2015lza}, the Atacama Cosmology Telescope (ACT) \cite{2014JCAP...04..014D}, and South Pole Telescope (SPT) \cite{0004-637X-743-1-90}, has led to first detections of secondary anisotropies, those imprinted in the CMB, following recombination, as the photons traverse large scale structures. This includes  the measurement of the thermal and kinetic Sunyaev-Zel'dovich effects (tSZ and kSZ, respectively)  \cite{1980MNRAS.190..413S}  and gravitational lensing of the CMB \cite{Hill:2013dxa}. 

Marked improvements in sensitivity, and an expansion of frequencies surveyed, at arcminute scales will be available with upgrades to the ACT and SPT facilities, Advanced ACTPol \cite{2010SPIE.7741E..1SN} and SPT-3G \footnote{https://pole.uchicago.edu}, and the construction of new facilities, including the Simons Observatory \footnote{http://simonsobservatory.org/}, the CCAT-prime observatory \footnote{https://www.ccatobservatory.org/} and  a next generation `Stage-4' ground-based CMB experiment, ``CMB-S4" \cite{Abazajian:2016yjj}. These promise a wealth of secondary anisotropy data  that could provide rich and mutually complementary information about the properties and evolution of galaxies and galaxy clusters, based on tracers of the ionized gas and gravitational potential.

This paper focuses on the impact of extraction and analysis assumptions on kSZ signal and covariance estimation. The kSZ effect is a Doppler distortion in the CMB  produced by the bulk motion of the cluster with respect to the CMB rest frame. Accurate measurements of the kSZ might therefore allow the inference of  peculiar motions of the most massive structures in the universe, and provide a powerful probe of the large scale structure (LSS) of the universe. The LSS growth rate can provide insights to central questions in cosmology, including the evolution of dark energy and cosmic modifications to gravity over cosmic time, and constraints on the sum of the neutrino masses \cite{Bull:2011wi, Mueller:2014nsa,Mueller:2014dba,Flender:2015btu}. 

While the tSZ  has a distinctive frequency dependence that can be used to extract it from multi-frequency measurements, the kSZ effect is frequency-independent and approximately twenty times weaker.  A variety  of techniques are being developed to extract the kSZ at cluster locations obtained from external LSS survey catalogs at other frequencies. Cluster bulk flows have been estimated with the kSZ effect \cite{Haehnelt:1995dg,Kashlinsky:2000xk, 2009ApJ...707L..42K} using WMAP and Planck data and X-ray detected clusters \cite{Kashlinsky:2009dw, Kashlinsky:2010ur, 2011ApJ...737...98O, 2012ApJ...758....4M,2014A&A...561A..97P}.
Combined kSZ and tSZ measurements of individual clusters have yielded constraints on their peculiar velocities \cite{1997ApJ...481...35H,2003ApJ...592..674B}. 

A number of recent surveys  have demonstrated the potential for extracting the pairwise kSZ signal from CMB data in combination  with galaxy surveys to locate the clusters \cite{Kashlinsky:2009dw, Kashlinsky:2000xk, 1538-4357-515-1-L1, 2013MNRAS.430.1617L, 2017JCAP...01..057S}. The first statistically significant kSZ detection was achieved using the pairwise estimator with ACT CMB data \cite{Hand:2012ui} in tandem with spectroscopic luminous red galaxies (LRGs) from the Baryon Oscillation Spectroscopic Survey (BOSS) survey \cite{Ahn:2013gms}. This has been extended upon with kSZ measurements from  SPT \cite{Soergel:2016mce} in combination with photometric galaxy survey information from the Dark Energy Survey (DES)  \cite{Rykoff:2016trm}, with Planck \cite{Ade:2015lza} in combination with the Sloan Digital Sky Survey (SDSS) Central Galaxy Catalog (CGC) \cite{Abazajian:2008wr}  and from ACTPol with BOSS \cite{Schaan:2015uaa,DeBernardis:2016pdv}. New techniques beyond the pairwise statistic have recently been developed to measure the kSZ signal, including new matched-filter estimators \cite{2014MNRAS.443.2311L} and three-point statistics \cite{Hill:2016dta, 2016PhRvD..94l3526F}.
  
Galaxies are often used as proxies to identify and locate the centers of galaxy clusters for kSZ extraction. A typical assumption is that the  brightest halo galaxy (BHG)  pinpoints the center of the cluster region where the temperature of the CMB photons is altered due to the motion of cluster's ionized gas. This approach follows the `Central Galaxy Paradigm' \cite{vandenBosch:2005} in which the central galaxy in the dark matter halo is the most massive and luminous galaxy due to continued gas accretion, relative to tidal stripping and ram-pressure quenching of star formation in captured satellite galaxies.  In observations, miscentering biases can arise from the misindentification of the cluster central galaxy. In spectroscopically selected galaxy catalogs, luminosity and color cuts are used to isolate the brightest, red galaxies. Photometrically selected catalogs can typically include many cluster members, or potential members, not just the single brightest, and have been used to determine how well BHGs trace the cluster center. 

In simulations the cluster central galaxy is defined as the location of the gravitational potential minimum, coinciding with the projected center of the electron distribution for kSZ extraction. Using the Red-sequence Matched-filter Probabilistic Percolation (redMaPPer) \cite{Rozo:2014sla}  algorithm studies have found $\sim$20\% \cite{Hoshino:2015zza} to $\sim$40\% \cite{Skibba:2010ez} of the BHGs are off-centered when considering ranked centering and cluster membership probabilities. As a result there can be detriments to the kSZ signal due to the fact that even the true BHG does not always trace the location of the potential minimum \cite{Flender:2015btu}. RedMaPPer has also provided evidence of anti-correlations between central galaxy brightness and cluster mass at fixed richness that could signal cluster mergers that might result in  galaxy position disruption that would effect the applicability of the central galaxy paradigm \cite{Hoshino:2015zza}. A variety of analytical offset models, based on observations, are also used to model and quantify the impact of miscentering for BHG data; they typically assume a fraction of galaxies have a Gaussian, or double Gaussian, miscentering distribution \cite{Johnston:2007uc, Saro:2015lqu}. Previous miscentering work has been done applying analytical offset models with simulations, finding a biased amplitude averaged over separation bins up to 200 Mpc as large as 11\% \cite{Flender:2015btu}.

In addition to transverse mis-identification of cluster positions,   photometric redshift errors  induce radial uncertainties in cluster locations  that can also dilute statistical power in the pairwise estimator \cite{Keisler:2012eg,Soergel:2016mce}. 

In this paper we characterize the size and nature of the impact of a variety of analysis assumptions for the extraction of the pairwise kSZ signal and covariance estimation for current CMB and LSS datasets. We consider different covariance estimation techniques, assumptions in kSZ decrement estimation, and the  impact of both transverse miscentering and photometric redshift (`photo-z') errors. We utilize data from  Planck and WMAP CMB surveys, and SDSS, WISE, and DECaLs LSS surveys. Understanding the impact of these assumptions on current surveys is an important practical step in order to assess the implications of the kSZ science potential of future spectroscopic and photometric surveys, including DESI, LSST and Euclid, and complements work in tandem on simulated datasets \cite{Keisler:2012eg,Flender:2015btu}. 

Our work is organized as follows: in section \ref{sec:data}, the CMB datasets and large scale structure surveys used in the analysis are described. The analytical formalisms used for: the pairwise estimator and background on signal extraction; covariance estimation; and miscentering models, are outlined in section \ref{sec:form}. Our results  are discussed in section \ref{sec:analysis}, and the findings and their implications for future work are drawn together in section \ref{sec:conclusions}.

\section{Datasets}
\label{sec:data}
In order to extract the kSZ signal temperatures, we cross-correlate  CMB maps and galaxy survey catalogs, from which the locations of the  brightest central galaxies are, as the default, assumed to trace the cluster centers. In this section we describe the different CMB and galaxy survey data sets used in the analysis.

\subsection{CMB maps}
\label{sec:data:cmb}
%
In this analysis we use the publicly available, foreground-cleaned \textit{Planck} {SEVEM} (Spectral Estimation Via Expectation Maximization) map \cite{Ade:2013crn}\footnote{http://pla.esac.esa.int/pla/}. The NSIDE = 2048  HEALPix  \footnote{http://healpix.sourceforge.net} SEVEM map covers, after confidence masks and foreground subtraction,  approximately 85\% of the full sky temperature map and a 5 arcmin FWHM \cite{Ade:2015lza}. We have also conducted comparable analyses with the the foreground-, dust-, and tSZ-cleaned {LGMCA} (``local-generalized morphological component analysis") map \cite{Bobin:2014mja}\footnote{http://www.cosmostat.org/product/{ LGMCA}\textunderscore cmb}, derived from a joint analysis of WMAP and Planck and find the results are nearly identical, with no significant differences induced for the aperture photometry as result  of the different foreground removal approaches, including for tSZ removal, used to produce the maps.
\begin{figure*}[!t]
	{
	\includegraphics[width=0.49\textwidth]{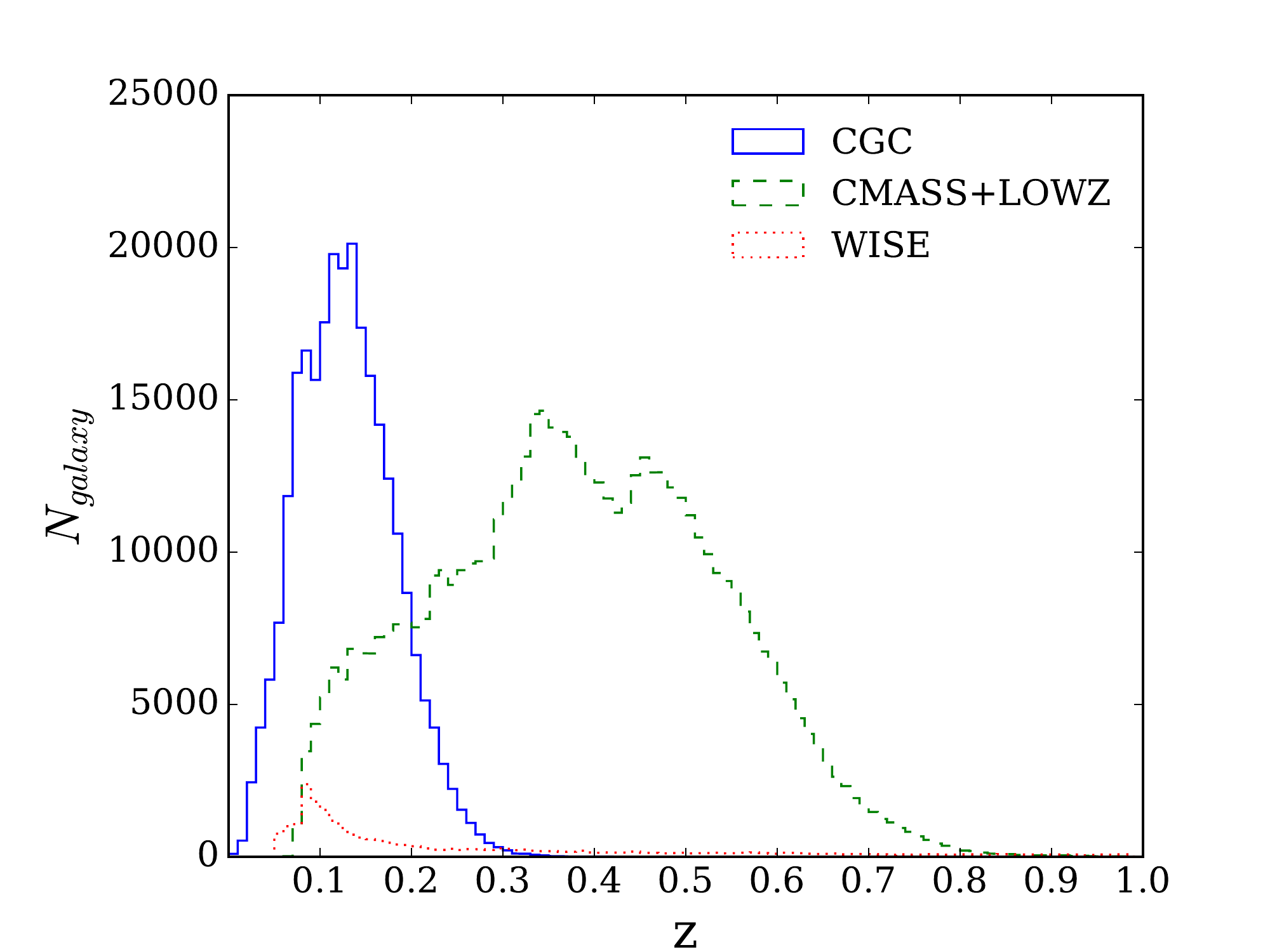}
	\includegraphics[width=0.49\textwidth]{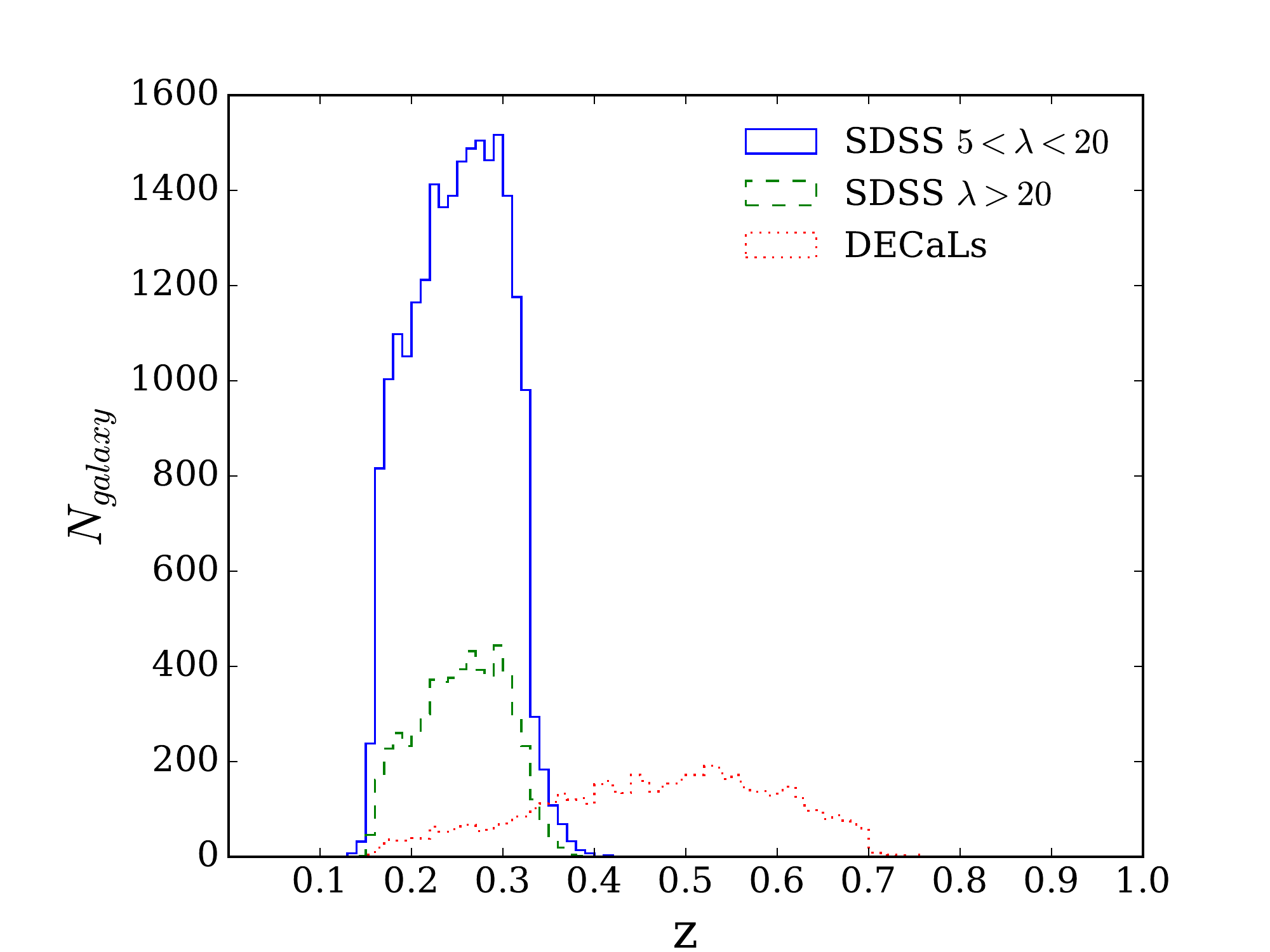}
	}
	\caption{[Left] The galaxy number distributions as a function of redshift, $z,$  for the spectroscopic-selected galaxy samples, showing SDSS CGC [full line],  CMASS+LOWZ [dashed line] and WISE [dotted line] samples. [Right] The redshift distributions for the redMaPPer catalogs:  two SDSS  redMaPPer samples with low [full] and high [dashed] richness, $\lambda$, and a sample from DECaLs [dotted]. Further details about the samples are given in Table \ref{tab1}.}
	
	\label{fig1}
\end{figure*}

\subsection{Galaxy samples}
\label{sec:data:lss}

Two methods with galaxy surveys are used to identify clusters and their center location to extract the kSZ from the CMB samples. 

The first approach uses spectroscopic LSS data, which gives precise redshifts for a select sample of bright, red galaxies, to identify and locate each cluster by targeting the brightest central galaxy following the Central Galaxy Paradigm. One option is to impose an aggressive luminosity cut, to include only the very brightest in the sample for cross-correlation. This avoids multiple galaxies being included in each cluster but also has the risk of not identifying all clusters, by excluding some of the less luminous brightest central galaxies. An alternative is a less aggressive luminosity cut, combined with the exclusion, around each bright galaxy, of other fainter galaxies  within  a characteristic cluster radius from the sample. This will lead to a more complete identification of cluster but, if the luminosity threshold is too low, could lead to satellites being mis-identified as the cluster center.

Three primarily spectroscopically selected ``Brightest Galaxy"  catalogs and three primarily photometrically selected redMaPPer galaxy samples are considered in this work. Their redshift distributions are shown given in Figure \ref{fig1}, and the selection criteria, redshift information and total catalog size for each sample are described below.

\subsubsection{Brightest Galaxy Catalog datasets}
\label{sec:data:lss:spec}

For the first of our brightest galaxy catalogs, the SDSS Central Galaxy Catalog (CGC), we utilize the catalog  from \cite{Ade:2015lza}\footnote{Wenting Wang \& Carlos Hern\'{a}ndez-Monteagudo, private communication}. This contains $262,671$ sources, selected based on the isolation criterion such that each galaxy is the brightest extinction-corrected $r$-band galaxy with $r<17.7$ found within 1.0 Mpc transverse distance and redshift difference corresponding to $1000$ km s$^{-1}$. It includes an additional cut to account for potential companions which fail to have a spectroscopic redshift due to fiber collision; this is done by cross-comparing using the photometric ``redshift-2" catalog \cite{2009MNRAS.396.2379C}, available at \footnote{http://das.sdss.org/va/photoz2/}. 
The luminosity range is $-29.8<R<-11.9$. Recent work in the ACT collaboration used a more aggressive luminosity cut corresponding to approximately $R<-22.8$ for their BOSS galaxy sample sample \footnote{http://data.sdss3.org/sas/dr11/boss/lss/}. To compare the impact of such variations in the luminosity cutoff, we also consider a stricter CGC sample with absolute $R$-band Petrosian magnitude, $R<-20.5$, corresponding to the mean value for the sample, compared to the initial $R<-11.9$, retaining $140,933$ galaxies of the original $262,671$. 

We create a WISE cluster-center catalog based on the color criterion used for the galaxy sample in \cite{2016arXiv160301608H} comprised of $10^6$ galaxies with $r<17.7$ obtained from the WISE All-Sky Data Release Catalog \cite{2013AJ....145...55Y}. Selection conditions are then applied: isolating the brightest extinction-corrected red galaxy within bins of roughly 12 by 12 arcmin across the sample, further removing any galaxies which are not the brightest in 1 Mpc transverse and radial separations, and cross-matching using the WISEx\textunderscore match function \footnote{https://skyserver.sdss.org/dr12/}, in the SDSS DR12 CasJobs query database to obtain redshifts, retaining $63,085$. Two additional cuts are made based on redshift: galaxies with $z<$ 0.05  are removed as we find the aperture photometry method presents at   redshifts below this threshold require unreasonably large apertures sizes to trace the kSZ signal. Roughly $90\%$ of the original sample redshifts are  photometric, we exclude approximately 20,000 low redshift galaxies  with   $\sigma_{phot}/z$ $>$ 0.1 to exclude those with poor photo-z estimates; this reduces the final sample size to $24,731$. 

We consider a combined CMASS and LOWZ sample from the public DR11 Large Scale Structure catalog \footnote{http://data.sdss3.org/sas/dr11/boss/lss/} from the SDSS-III Baryon Oscillation Spectroscopic Survey (BOSS) \cite{2013AJ....145...10D}. To the  combined base catalog, exceeding one million galaxies, we apply a magnitude criteria corresponding to galaxies with absolute $R$-band Petrosian magnitudes brighter than $-21.4$ in order to maintain a sufficiently selective sample of bright galaxies, giving a sample of 555,307. The choice of $R<-21.4$ corresponds to the mean value for the full selection of galaxies in the initial catalog.

\subsubsection{redMaPPer datasets}
\label{sec:data:lss:phot}

In addition to the principally spectroscopic BCG catalogs described above, we consider three catalogs created using the redMaPPer algorithm, which include a higher fraction of photometric redshifts \cite{Rykoff:2013ovv}. These allow us to simultaneously study the relative impacts of photometric redshift and transverse miscentering uncertainties related to using the BHG to pinpoint the cluster center. 

The photometric redMaPPer (RM) data identifies and locates galaxy clusters using iterative red-sequence modeling, based on the fact that old, red galaxies make up the bulk of clusters and that the brightest, most massive galaxies reside in the cluster center.  With the redMaPPer algorithm, a membership probability $P_{mem,ij}$ is assigned to the $i^{th}$ galaxy associated with the $j^{th}$ cluster and also assigned a rank in centering probability in that cluster, $P_{cen,ij}$ \cite{Rykoff:2016trm}. 

We consider  cluster catalogs selected based on their ``richness", $\lambda$, a reasonable measure of cluster mass for  photometric surveys \cite{Rozo:2015mmv}. This is determined by sum of the membership probabilities, for all  $N_{gal}$ galaxies associated with the cluster, 
\bea
\lambda_j =\sum_{i=1}^{N_{gal}} P_{free,i} P_{mem,ij}, 
\eea  
where $P_{free}$, typically $\approx 1$, is the probability that a galaxy within the cluster is not  a member of another cluster within a characteristic, richness dependent cutoff radius, 
\begin{eqnarray}
 R_c(\lambda) = R_{0}\left(\frac{\lambda}{100}\right)^\beta,
 \label{eq:Rc}
\end{eqnarray}
with $R_{0}$ = 1.0Mpc$/h$ and $\beta$ = 0.2 \cite{Rykoff:2012vja}.

We use the redMaPPer central galaxy (RMCG) catalog from the SDSS DR8 cluster catalog (v5.10) \cite{2016ApJS..224....1R} and the DECaLs DR2 catalog v6.4.12 \cite{Rozo:2015mmv,Rykoff:2016trm}. The central galaxy is found by selecting the galaxy within each cluster with the highest centering probability as computed with the redMaPPer cluster-finding algorithm \cite{Rykoff:2013ovv}.

The SDSS RMCG high-richness catalog, with $\lambda > 20$,  includes $7,730$ SDSS DR8 clusters (of which 5,818 center-proxy galaxies meet the criteria to be an LRG) and $507,874$ total member galaxies spanning $0.16 < z < 0.33$. Brightness is restricted to $i < 21.0$. 

We also analyze a subset over the same redshift range for which $5 < \lambda < 20$ for which there are $22,492$ cluster-center proxy LRGs and $1,878,746$ total member galaxies  \cite{Hoshino}, \cite{Rykoff:2016trm}. However, we recognize that this dataset is not as reliable as high richness sample, due to the lower number of member galaxies and the lower likelihood of those galaxies being associated with clusters.

The final catalog we consider is not publicly available \footnote{Eduardo Rozo \& Eli Rykoff, private communication}, but is a preliminary redMaPPer central galaxy catalog of the Dark Energy Camera Legacy Survey (DECaLs) DR2 RMCG  that overlaps with the CMASS+LOWZ sample. 
The higher redshift DECaLs RMCG  contains $5,870$ most likely cluster-center proxies,  out of 30,020 total member galaxies, of which 43\% have photometric redshifts only. All have $\sigma_z/z<0.1$, eliminating the need apply this criteria as we did for WISE.

\section{Formalism}
\label{sec:form}

\subsection{Pairwise estimator}
\label{sec:form:pw}

The temperature distortion in the CMB induced by the cluster's peculiar motion is given by \cite{1970Ap&SS73S},
\begin{equation} 
\frac{\delta T_{kSZ}}{T_0}(\hat{r}) = -\int dl\,\sigma_T \, n_e  \frac{{\bf v}\cdot {\bf\hat r}  }{c} 
\end{equation} 
where $n_e$ is the electron number density and $\sigma_T$ is the Thomson cross-section. A positive peculiar velocity, {\bf v}, relates to motion away from the observer, so induces a negative kSZ effect. For the case in which the kSZ signal is dominated by a single cluster along the line of sight 
\bea
\frac{\delta T_{kSZ}}{T_0}(\hat{r})=-\tau\frac{v_r}{c}
\eea
where $\tau$ is the cluster optical depth and $v_{r}$ is cluster line of sight peculiar velocity.  

The kSZ effect itself is  a direct measure of the cluster momentum, since it is dependent on both the velocity of the cluster and the number density of electrons. 
As a measure of their gravitational infall,  clusters are likely to be moving towards each other, and this should show up in the correlation of cluster velocities, and hence in the related kSZ signature in the CMB. To obtain the peculiar velocity correlations traced by the kSZ effect, we employ the pairwise estimator,  derived by Ferreira et al. \cite{1538-4357-515-1-L1}. 

The pairwise momentum estimator is given by \cite{Hand:2012ui}, 
\begin{equation} 
\hat p_{kSZ}(r) = - \frac{\sum_{i<j} (\delta T_i - \delta T_j) c_{ij}}{\sum_{i<j} c^2_{ij}}, 
\label{eq:phat}
\end{equation} 
where the sum is over all galaxy cluster pairs, located at positions ${\bf r}_i=\{{ \bf\hat{r}}_i, z_i\}$ and  ${\bf r}_j=\{{\bf\hat{r}}_j, z_j\}$, separated by a distance $r=|{\bf r}_{ij}|=|{\bf r}_i-{\bf r}_j|$, and with ${\bf\hat r}$ the unit vector in the direction of {\bf r}. $\delta T_i$ represents the relative kSZ temperature at the $i^{th}$ cluster location. The weights $c_{ij}$ are given by
\begin{equation} 
c_{ij} =  \hat r_{ij} \cdot \frac{\hat r_i + \hat r_j}{2} = \frac{(r_i - r_j)(1+cos\alpha)}{2\sqrt{r_i^2 + r_j^2 - 2r_i r_j cos\alpha}}
\end{equation} 
where $\alpha$ is the angle between $\hat r_i$ and $\hat r_j$.

 \begin{figure}[t!]
 	{	\includegraphics[width=0.50\textwidth]{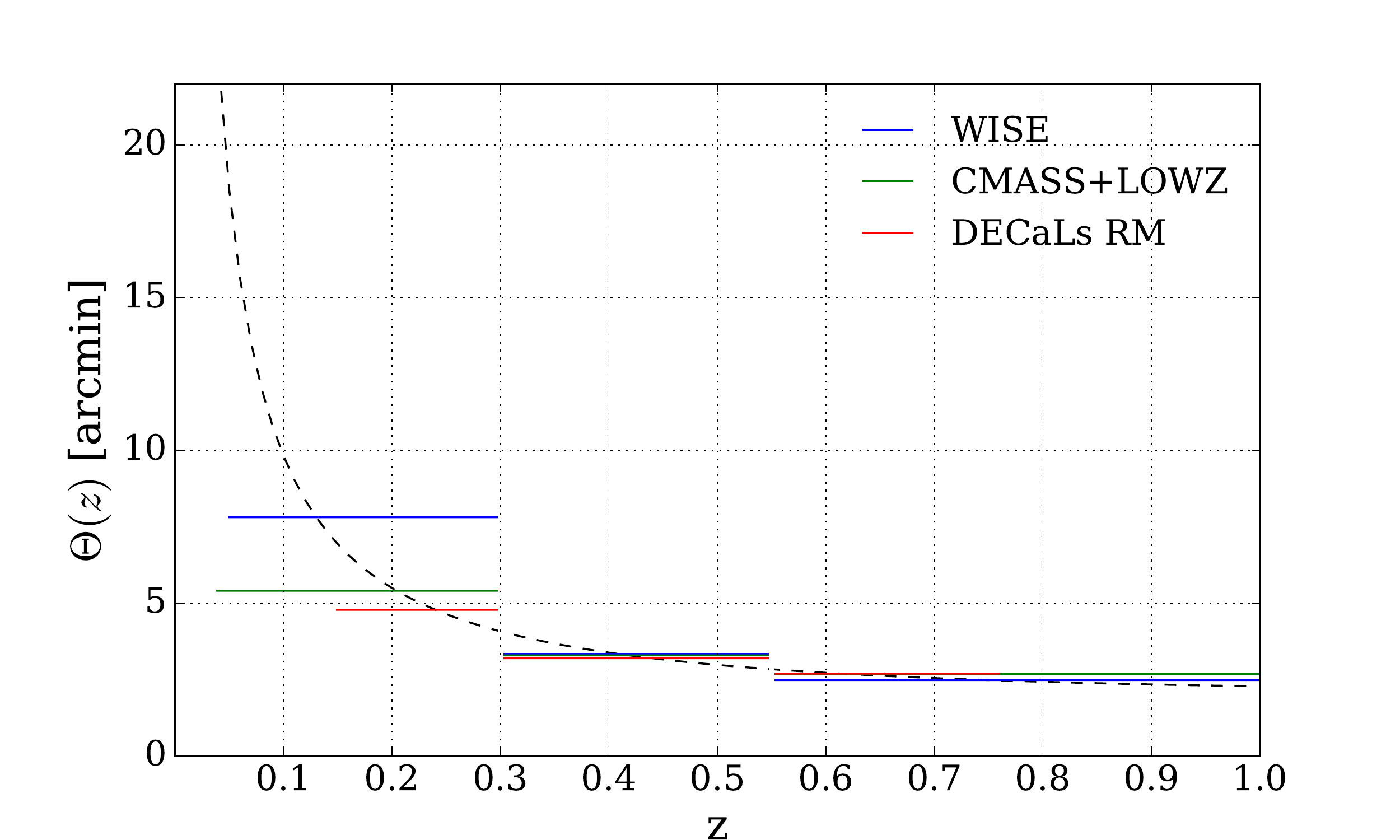}	
 	}
 	\caption{The redshift-dependent aperture used for the aperture photometry kSZ temperature decrement estimation  for the WISE [blue] and CMASS+LOWZ [green] samples and DECaLs redMaPPer (RM) catalog [red], which have galaxies distributed over extended redshift ranges.  The  angular size of a 1.1 Mpc galaxy cluster as a function of the redshift is shown for comparison [dotted line].}
 	\label{fig2}
 \end{figure}

\subsection{Signal extraction}
\label{sec:form:sigex}
We employ aperture photometry (AP)  to isolate  the kSZ signal from the CMB. This technique relies upon the kSZ being localized in the cluster, while the primordial CMB modes, correlated over longer wavelengths, are removed by differencing the cluster region with an annular region immediately adjacent to it.  The aperture photometry temperature, $T_{AP} = \langle T_{inner} \rangle - \langle T_{annul} \rangle$,  is the difference between the average CMB pixel temperature within a given angular radius $\Theta$, comparable to the cluster size, to that within an  annulus, outside the radius, of width $\sqrt{2}\Theta$. A cluster of scale 1.1 Mpc is used as the basis for the typical angular size for the aperture. 

As summarized in Figure \ref{fig1}, our analyses include some galaxy samples that are relatively compact in redshift space, SDSS CGC, and the SDSS redMaPPer datasets, and others that have broader redshift distributions, WISE and CMASS+LOWZ. To accommodate these differences, we compare pairwise results using two alternative aperture size criteria. For all datasets, we consider a single, redshift-independent aperture, fixed by the typical cluster angular size, $\Theta(\bar{z})$ at the survey's mean redshift, $\bar{z}$. For  the datasets that are extended in redshift, a redshift-dependent aperture, $\Theta(z)$, which is  binned for the clusters in each sample in three ranges, $z<0.3$, $0.3<z<0.55$, and $z>0.55$ is also considered. The values of the redshift-dependent aperture are shown in Figure \ref{fig2} and the single aperture values for each sample summarized in Table \ref{tab1}.

The cluster kSZ decrement is  then given by
\begin{equation} 
\delta T_i(\hat{r}_i, z_i, \sigma_z, \Theta) = T_{AP}(\hat r_i,\Theta) - \bar{T}_{AP}(\hat r_i, z_i, \sigma_z,\Theta). 
\label{eq:Ti}
\end{equation} 
 where $T_{AP}(\hat r_i, \Theta)$ corresponds to the kSZ amplitude
 estimate obtained at the angular position of the $i^{th}$ galaxy with an aperture size $\Theta$, and $\bar{T}_{AP}$ is the averaged aperture temperature  over all cluster locations within a Gaussian distributed redshift range centered on the cluster, $z_i$, with width $\sigma_z$, introduced by \cite{Hand:2012ui} to account for possible redshift evolution of the tSZ signal in the sources: 
\begin{equation}
\bar{T}_{AP}(\hat{r}_i, z_i, \sigma_z,\Theta) = \frac{\sum_{j} T_{AP}(\hat{r}_i,\Theta) \exp\left(- \frac{(z_i - z_j)^2}{2\sigma_z^2}\right)}{  \sum_{j} \exp\left(- \frac{(z_i - z_j)^2}{2 \sigma_z^2}\right)}.
\end{equation} 
The sum is over all galaxies $j\neq i$ in the same redshift bin, and we take $\sigma_z = 0.01$ as in previous work by the Planck team \cite{Ade:2015lza}. 

  \begin{table}[t!]
  	\begin{tabular}{ | l | C{4.5em} | C{3.75em} | C{3.57em} | C{3.75em} |}
  		 \multicolumn{5}{l}{\it Brightest Galaxy Catalogs: }
		 \\ \hline    
		  { Sample} & $N_{gal}$ &  $\bar{z}$ & $\sigma_z$ & $\Theta(\bar{z})$ 
		 \\ \hline		
  		\hspace{0.1cm}SDSS CGC & 262,671 & 0.13 & 0.05 & 8.0'
  		\\ \hline
  		\hspace{0.1cm}WISE & 24,731  & 0.27 & 0.24 & 4.8'
  		\\ \hline 
  		\hspace{0.1cm}CMASS+LOWZ & 555,307  & 0.46 & 0.15 & 3.2'
  		\\ \hline
  	      \multicolumn{5}{l}{}
				 \\
  		\multicolumn{5}{l}{\it redMaPPer-derived Catalogs:}   
				 \\ \hline    
		  { Sample} &$N_{gal}$ &  $\bar{z}$ & $\sigma_z$ & $\Theta(\bar{z})$ 		
		\\ \hline
  		\hspace{0.1cm}SDSS ($5<\lambda<20$) &  \multicolumn{1}{c|}{22,492} & 0.25 & 0.05 & 4.7'
  		\\ \hline
  		\hspace{0.1cm}SDSS ($\lambda>20$) & \multicolumn{1}{c|}{5,818} & 0.26 & 0.05 & 4.6'
  		\\ \hline 
  		\hspace{0.1cm}DECaLs &  \multicolumn{1}{c|}{5,870} & 0.47 & 0.13 & 3.1'
  		\\\hline
  	\end{tabular}
  	\caption{Overview of the galaxy samples considered in the analysis: [upper] three spectroscopically selected brightest galaxy samples and [lower] three primarily photometric redMaPPer Central Galaxy Catalogs (RMCG), for which the SDSS samples are delineated on the basis of richness, $\lambda$. For each sample, the number of galaxies in the same, $N_{gal}$,  their mean redshift, $\bar{z}$,  and standard deviation, $\sigma_z$, and the aperture size of a $1$ Mpc scale at $\bar{z}$, $\Theta(\bar{z})$, is given.}  
  	\label{tab1}
  \end{table}

\subsection{Covariance estimation}
\label{sec:form:cov}

We compare covariance estimates using two distinct methods that have been used in pairwise kSZ studies in the literature: CMB map rotations and jackknife (JK) resampling. 

For the angular rotation method, as in \cite{Ade:2015lza}, we produce 50 sets of angular rotations relative to the real cluster-center positions, using a displacement step of three times the aperture radius adopted, and have confirmed that there is very little deviation in the variance with the number of rotations performed. The assumption is that  the displaced samples  reflect  a set of null realizations. 

For the JK covariance estimate, we  create resamples of the pairwise kSZ measurement by binning the clusters into $N_{JK}$ subsamples, removing one, and then computing the pairwise estimator according to the remaining $(N_{JK}-1)$ subsamples. This is done such that each subsample is removed precisely once. The covariance matrix is then given by \begin{equation} \hat C_{ij}^{JK} = \frac{N_{JK}-1}{N_{JK}}\sum_{\beta=1}^{N_{JK}} (\hat p_i^\beta - \bar p_i) (\hat p_j^\beta - \bar p_j),
\end{equation} where $\hat p_i^\beta$ is the pairwise kSZ signal in separation bin $i$ and JK subsample $\beta$, with mean of the $N_{JK}$ samples, $\bar p_i$ \cite{2016MNRAS.461.3172S}. We use $N_{JK}=100$ submaps for all analyses and find that covariance estimates  are largely insensitive to changes around this subsample size (we considered $N_{JK}=$ 50, 100, 250). We also confirmed that the results were unaffected by limiting to longitudinal rotations only, as in \cite{Ade:2015lza}, versus including both longitudinal and latitudinal as we use in modeling the miscentering offset.

\subsection{Cluster centering estimation}
\label{sec:form:center}

The effectiveness of the aperture photometry technique is dependent on the ability to identify and locate the center of each galaxy cluster. As discussed in the introduction, astrophysical processes including cluster merges can introduce systematic offsets in the locations of the brightest, most massive galaxies, typically expected to exist in the cluster's central region \cite{0004-637X-786-2-79}. 

We consider two  approaches to study the effect of cluster miscentering on the kSZ signal: 1) contrasting  redMaPPer selected catalogs of the brightest versus the most likely central galaxies, 2) using an analytic Johnston model (based on photometric catalog analyses) \cite{Johnston:2007uc}.

We directly test Central Galaxy Paradigm with the redMaPPer data by considering the differences in the predicted signal if the ``cluster center" is assigned to  RM-identified galaxies, other than the one with highest probability central galaxy. Following the approach discussed in  \cite{Hoshino:2015zza}, we create  two catalogs from the RM data, for each of two populations, $5>\lambda>20$ and $\lambda<20$: a redMaPPer central galaxy (RMCG) catalog, based on the highest rank cluster center, and the redMaPPer brightest galaxy (RMBG) catalog. 

For  the brightest galaxy catalogs, we then consider an analytical offset model by Johnston et al. \cite{Johnston:2007uc},  obtained through comparing weak lensing profiles with brightest galaxy positions in the SDSS galaxy clusters. It assumes that a fraction of the cluster sample, $f_J$, has precisely known centers, corresponding to where the brightest galaxy is close to the cluster's gravitational potential minimum, and the remaining $(1-f_J)$ have a brightest galaxy a distance $\doff$ from the center, with a probability following a Rayleigh distribution function of width $\sigma_J$: 
  \begin{equation}
 p_J(\doff) =  \frac{\doff}{\sigma_J^2}\exp\left(-\frac{\doff^2}{2\sigma_J^2}\right).
  \end{equation} 
  Johnston et al. find a reasonable fit to their BHGs with $\sigma_J$=0.42 Mpc$/h$, and find a richness dependent fraction of BHGs are well-centered, ranging from $f_J\sim$ 60$-$90\% for clusters of mass $5\times 10^{12} - 5\times 10^{14}M_{sun}/h$ respectively.
    
We estimate the fraction of clusters for which the brightest target galaxy is at the cluster center and the offset size for those that are miscentered, $f_J$ and $\sigma_J$, by using the redMaPPer samples, summing over  the probabilities of scenarios in which the galaxy with the highest centering probabilities may in fact not be a true member of the cluster, presented here in a slightly modified form from the centering probability in \cite{Hoshino}:
\begin{equation}
\begin{split}
 N \cdot f_J \approx & \sum_{i=1 BG} P_{cen,i}(P_{mem}^{1BG})  + \sum_{i=2BG} P_{cen,i}(1-P_{mem}^{1BG}) \ + \\
 & \sum_{i=3BG} P_{cen,i}(P_{mem}^{1BG})(1-P_{mem}^{2BG}) \ + \\
 & \sum_{i=4BG} P_{cen,i}(P_{mem}^{1BG})(1-P_{mem}^{2BG})(1-P_{mem}^{3BG}) + ...
 \end{split}
 \label{eq:fJ}
\end{equation} 
where N is the number of clusters. For each central candidate target (e.g. LRG) galaxy $i$, we calculate $P_{cen,i}$ weighted by the probability that it is the brightest member galaxy as well. Here $n$BG refers to the $n$th brightest target galaxy in the cluster. For example, when the brightest galaxy (1BG) in the cluster is also the central candidate galaxy, $f_J$ is found by summing the centering probabilities with just the first term. The centering probabilities are normalized such that $\sum_i P_{cen,ij}$=1. We consider the first five BG terms only because the probability that none of the first five brightest galaxies are cluster members is negligible. 

\section{Analysis}
\label{sec:analysis}


\subsection{Covariance method comparison}
\label{sec:analysis:cov}
 
 As described in section \ref{sec:form:cov}, the JK method involves splitting the dataset into some number $N_{JK}$ submaps and finding the variance based on computing the signal with each submap removed exactly once. The rotation method estimates the covariance using the variance between the pairwise signals obtained after random longitudinal displacements of cluster locations by three times the aperture size. 
 
\begin{figure*}[!t]
	{	
		\includegraphics[width=0.325\textwidth]{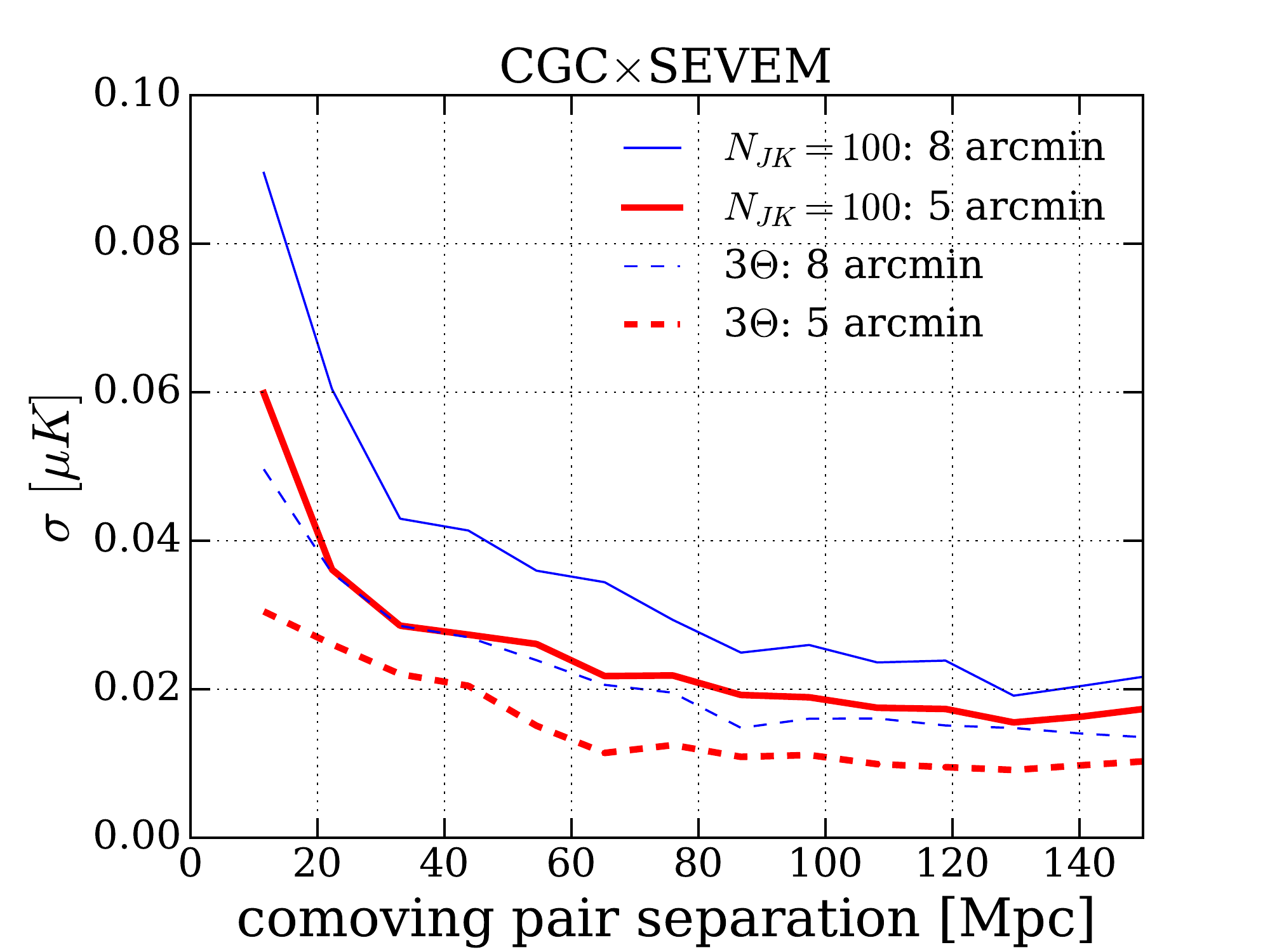}
		\includegraphics[width=0.325\textwidth]{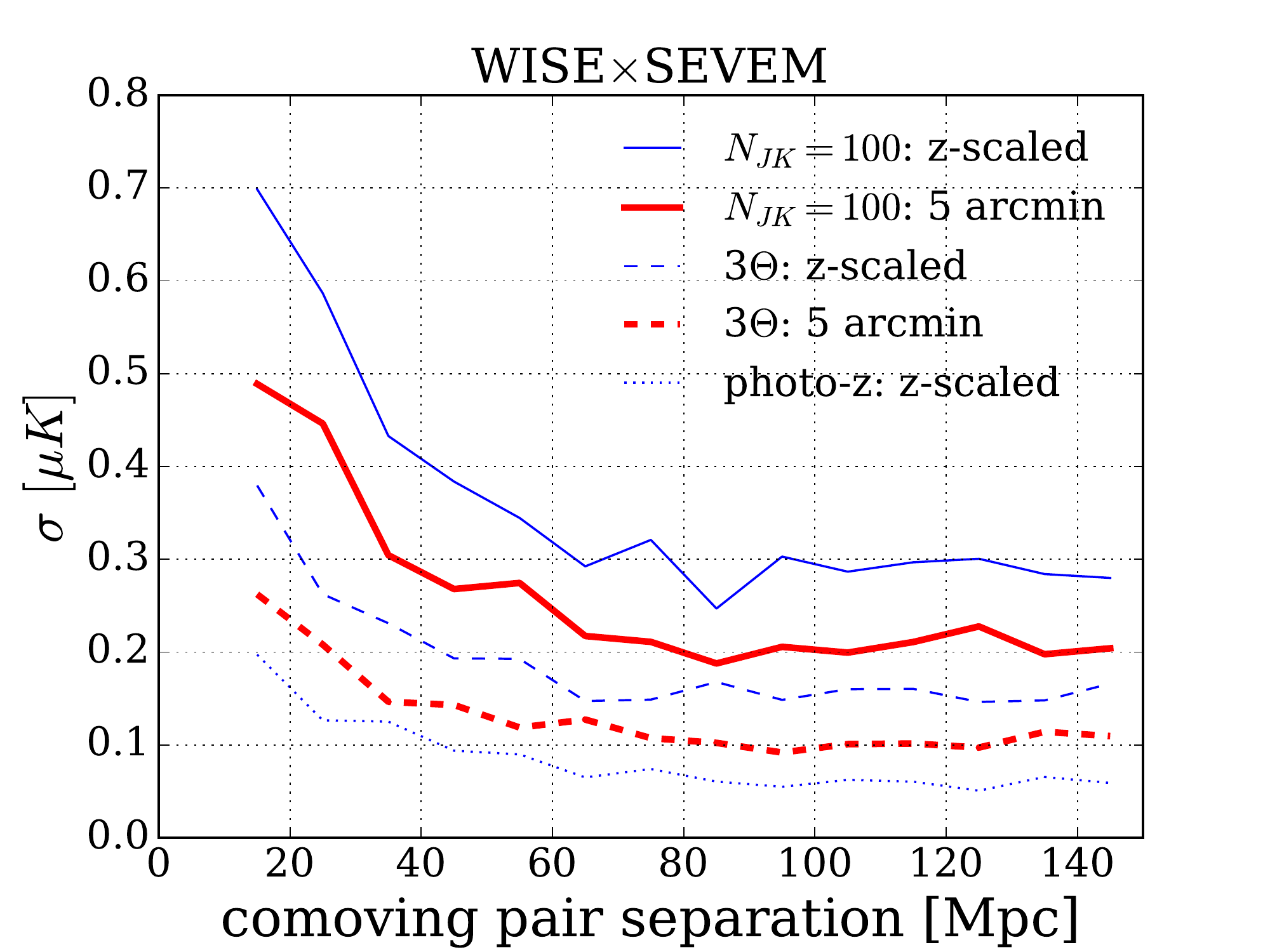}
		\includegraphics[width=0.325\textwidth]{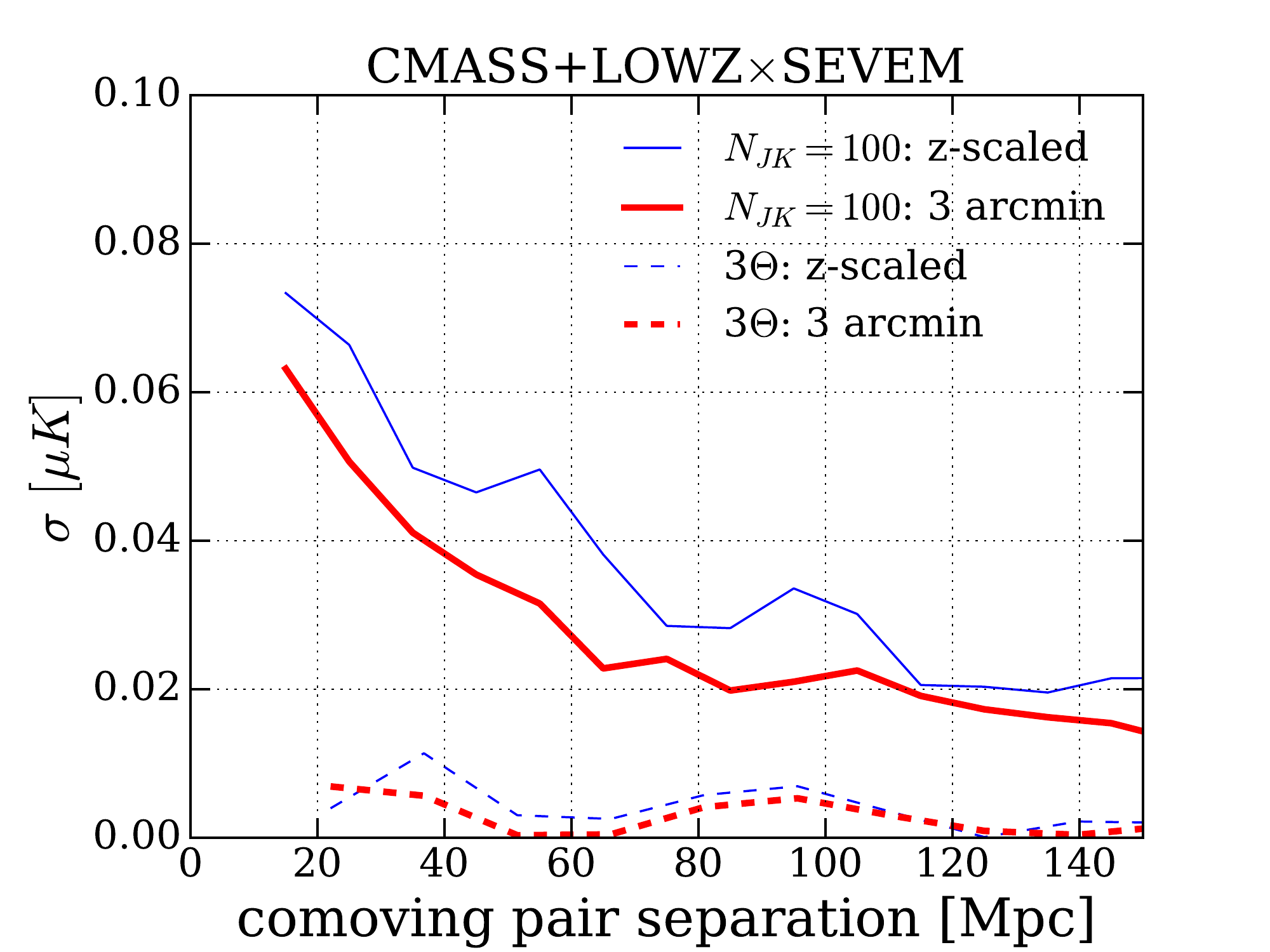}
	}
	\caption{Comparison of the standard deviation, $\sigma$, in $\hat{p}_{kSZ}$ for the $Planck$ SEVEM dataset as a function of comoving pair separation when cross-correlated with the SDSS CGC [left], WISE [center], and CMASS+LOWZ [right] galaxy samples using different covariance estimation techniques. In each plot, the errors estimated from a jackknife (JK) resampling, using $N_{JK}=100$ subsamples [full lines], are compared with those from maps  with 50 randomly oriented rotations of displacements of 3$\times$the aperture size, denoted by ``3$\Theta$"  [dashed lines]. The CGC$\times$SEVEM rotation errors with 5 [red, thick] and 8 [blue, thin] arcmin apertures, are consistent with those presented by the Planck collaboration \cite{Ade:2015lza}. The WISE and CMASS+LOWZ results compare statistical errors for fixed apertures [red, thick] with redshift dependent (`z-scaled') apertures [blue, thin].}
	\label{fig3}
\end{figure*}
  
In Figure \ref{fig3}, we show the results using the two covariance methods for our datasets with the \textit{Planck} SEVEM maps. We find that the JK method predicts larger covariances than the rotation method for the same aperture at all pairwise separations considered ($15-150$ Mpc). 

The covariance for both techniques is found to be sensitive to the aperture size.  At an intermediate separation of 55 Mpc, cross-correlations with the CGC catalog, the JK method estimates a standard error, $\sigma$, that is 1.7 larger than that predicted for rotations for an 8 arcminute aperture. For a smaller aperture size, that likely slightly underestimates the cluster size, the variance estimate is also smaller for both methods consistent with some of the signal being removed in the aperture photometry differencing. The WISE sample shows similar findings in comparing the two methods, with a factor of 2.3 difference in variance between the JK and rotation methods. It also shows that the aperture size choice when choosing either a fixed aperture at an average value or varying with redshift can affect the covariance estimate  for samples which are extended in redshift space. We find a large disparity between the two methods for  CMASS+LOWZ$\times$SEVEM, the cross-correlation with the most galaxies, with  factors of $\sim8$ between the JKs and rotations for both the redshift dependent and fixed apertures. 
 
Given the JK errors are more conservative than those from the rotations, and that the JK method is found to be reliable compared to estimates from simulations \cite{DeBernardis:2016pdv}, we use these errors for the analysis in the remainder of the paper. 

\subsection{$\hat p_{kSZ}$ sensitivity to galaxy sample, CMB map creation, and aperture choice}
\label{sec:analysis:base}
In this section we consider the impact on the pairwise estimator of assumptions that go into: the galaxy proxy sample selection; the CMB map generation; and the aperture photometry method.

\begin{figure*}[!t]
  	{
  		\includegraphics[width=0.49\textwidth]{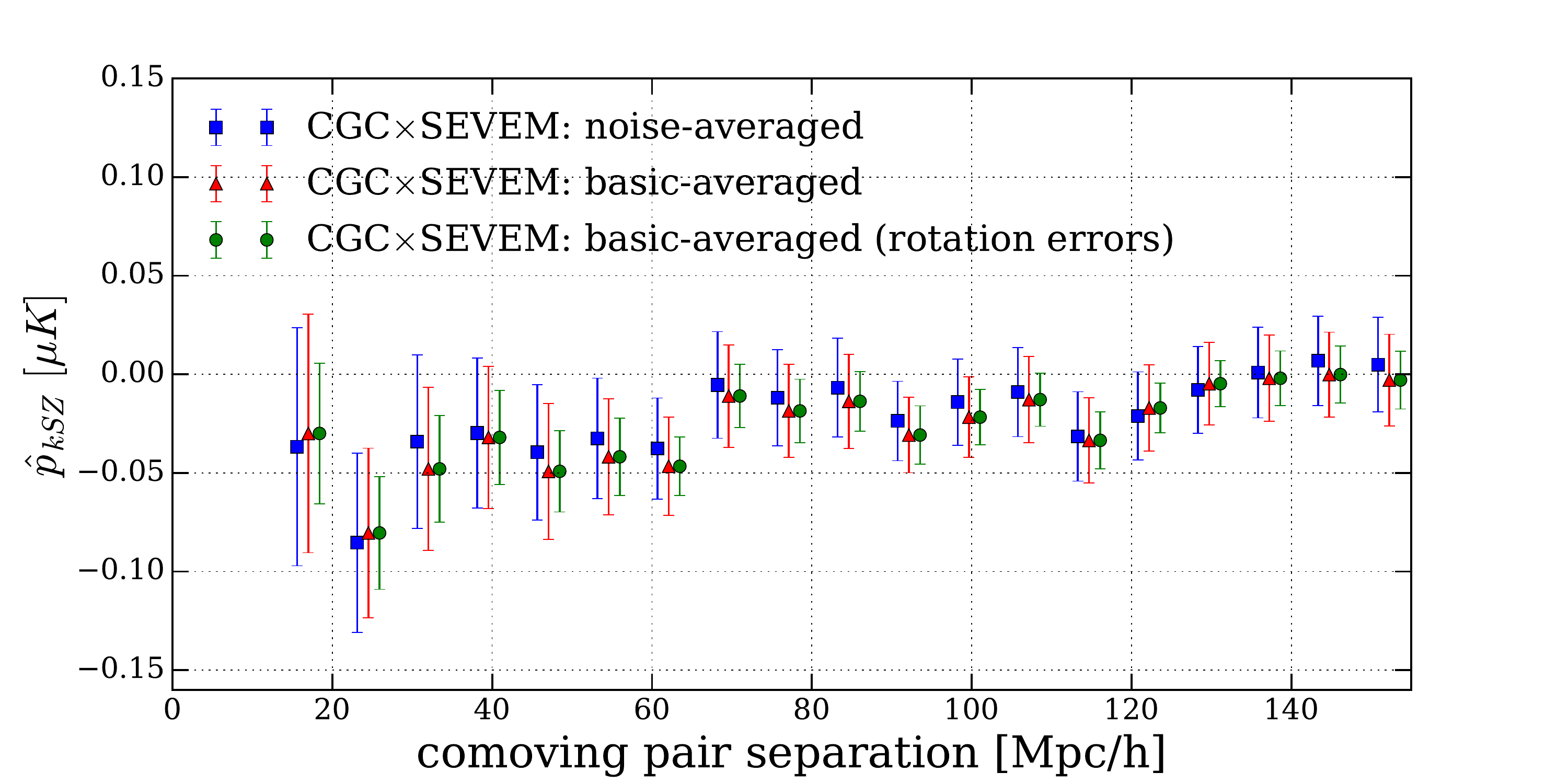}
		  \includegraphics[width=0.49\textwidth]{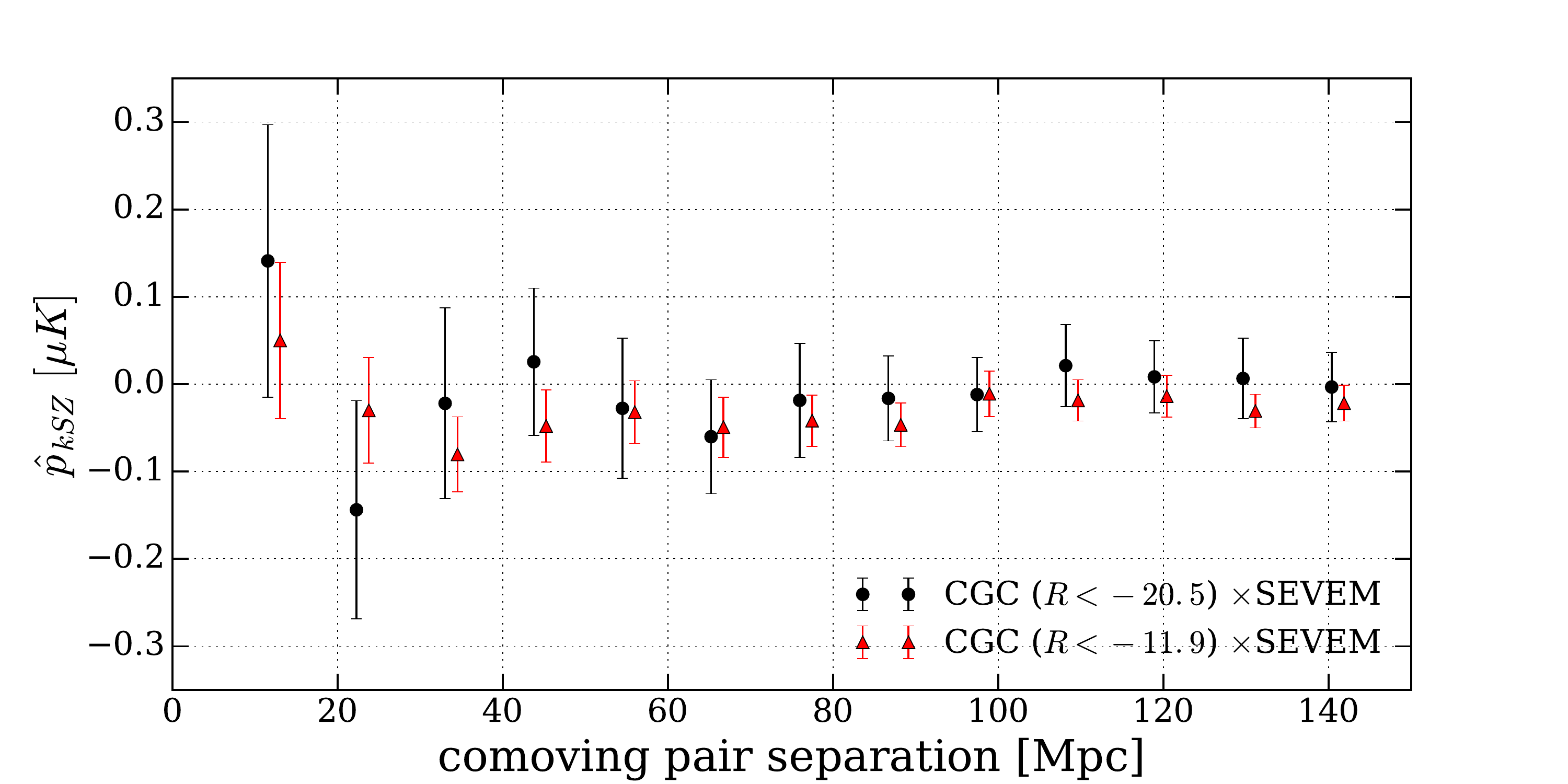}
  	}
  	\caption{
Comparisons of the $\hat p_{kSZ}$ signal and variance obtained for CGC$\times$SEVEM using an 8 arcmin aperture.
Shown in both plots  [red triangles] is $\hat p_{kSZ}$ using flat-weighing for the pixels, and JK errors for a galaxy sample, such that the absolute $R$-band magnitude falls within $-28.9<R<-11.9$. [Left]  A comparison of results using different CMB map weighting schemes and covariance estimators, comparing pixels that are noise-weighted [blue square] and flat-weighting [red triangles] with JK error estimates, and flat-weighted pixels with rotation errors, as in \cite{Ade:2015lza} [green circles]. To aid comparison with the Planck results, on this plot alone, we express separations in Mpc$/h$, but use Mpc in all future figures. [Right]  A comparison of different luminosity cut-offs for the CGC galaxy sample, such that the absolute $R$-band magnitude falls within $-28.9<R<-11.9$ [red triangles] and $-28.9<R<-20.5$ [black circles]. In both cases, only the brightest galaxy within a 1 Mpc radius is retained to isolate the best central galaxy candidate within a cluster volume.}
  	\label{fig4}
  \end{figure*}

In the left panel of Figure \ref{fig4}, we compare pairwise correlations for the CGC sample using Planck CMB maps with the same aperture photometry but in which the CMB pixels are noise weighted while in the other a flat-weighting is used. We find little difference, with only a marginal improvement in the $\hat p_{kSZ}$ aperture photometry results by down-weighting the noisy pixels in the map. Based on this we use a flat weighting in the remainder of the analyses.

In section \ref{sec:data:lss}, we describe the criteria used to develop the galaxy samples used as proxies to identify and locate clusters. These assumptions can be highly varied  across the analyses in the literature.

The right panel of Figure~\ref{fig4}, demonstrates the impact of the luminosity cut assumptions in the galaxy proxy catalog. $p_{kSZ}$ is shown for the SEVEM map when cross-correlated with the CGC catalog from the Planck analysis \cite{Ade:2015lza} an apparent  $r$-band magnitude cut was imposed, and with the redshift distribution of the sample, this translates into a redshift dependent absolute magnitude cut of $-29.8<R<-11.9$, comprising $262,671$ galaxies. We also show results for a catalog with a stricter luminosity cut,  $R<-20.5$, with $140,933$ galaxies. These magnitude cuts respectively correspond to luminosity thresholds of $\sim$3.5 $\times$ $10^6 L_{sun}$  and $\sim$8.8 $\times$ $10^8 L_{sun}$.  By comparison, recent work by the ACT collaboration considered a luminosity cut of $L > 7.9 \times 10^{10}L_{sun}$ \cite{DeBernardis:2016pdv}. The more conservative threshold was chosen so as to maintain at least 100,000 galaxies in the sample most likely to represent a central galaxy sample, while reducing potential contamination from  satellite galaxies.  The stricter selection criteria, especially at lower separations,  shifts $p_{kSZ}$ by more than the $1\sigma$ relative to that using the conservative criteria. The statistical uncertainties are also increased, consistent with the more aggressive cut decreasing the sample size. At 43 Mpc separation, for example, the errors differ of the stricter cut are larger than the initial sample by a factor of 2.0, while the signal is increased by a factor of 0.5.

\begin{figure*}[!t]
    	{
    		\includegraphics[width=0.49\textwidth]{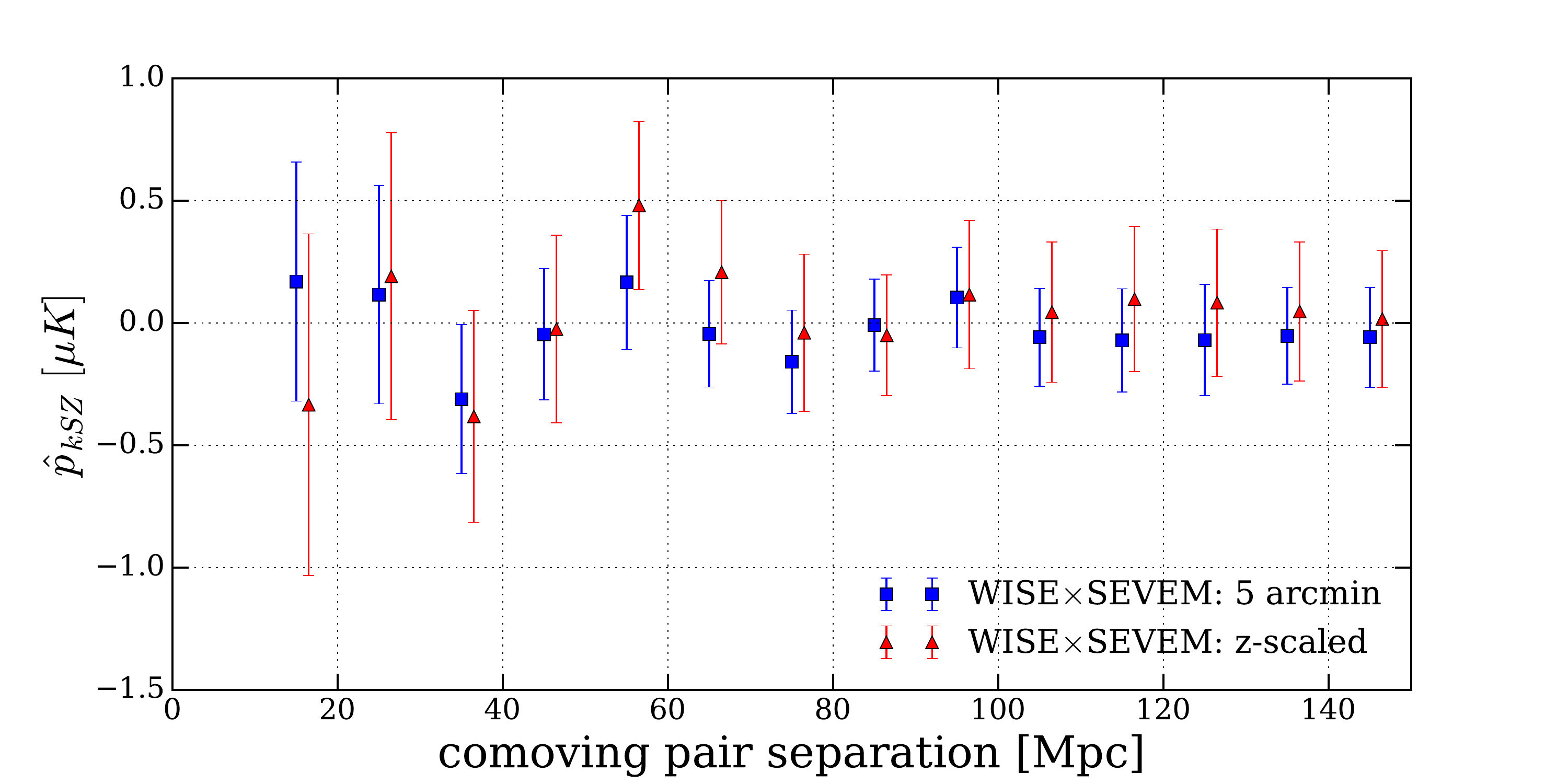}	
    		\includegraphics[width=0.49\textwidth]{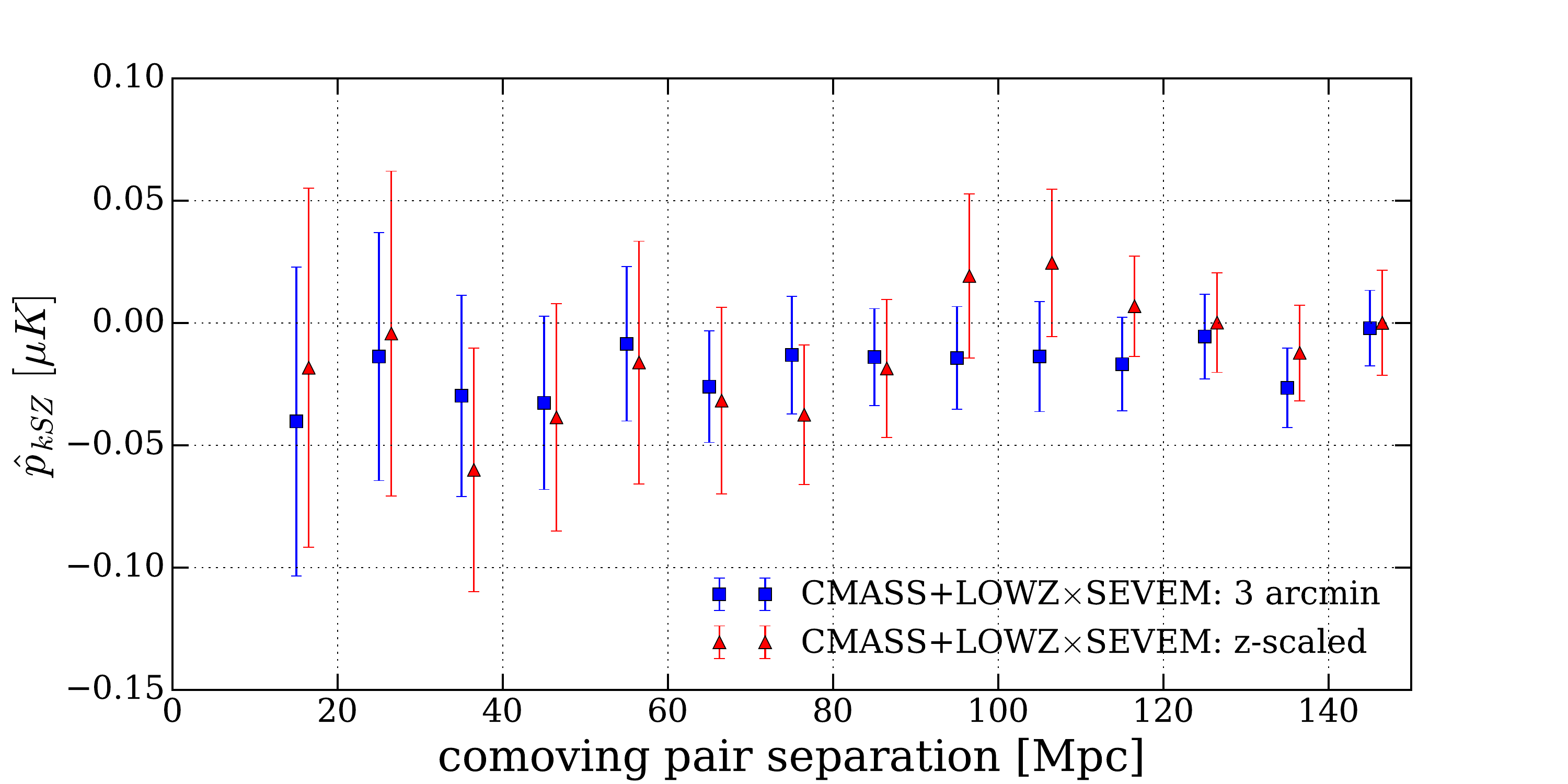}
    		
    	}
    	\caption{Comparisons of the $\hat p_{kSZ}$ signal and variance obtained for the SEVEM CMB data and   [left] the WISE  and [right] CMASS+LOWZ galaxy samples. In each  a  redshift dependent [blue square] and $z$-independent [red triangle] aperture, that closely matches the expected angular size of a cluster at each sample's mean redshift. }
    	
    	\label{fig5}
    \end{figure*}

In Figure \ref{fig5}, we study the robustness of the aperture photometry method to assumptions on aperture size and foreground contamination. We consider fixed angular apertures to range from 3 to 8 arcmins, based on the expectation scaled by the cluster size, $\Theta(\bar z)$, as given in \ref{tab1}. We also consider the redshift-scaled aperture for the WISE and CMASS+LOWZ datasets. Cases where the aperture is incorrectly selected will result in scenarios in which cluster data is being inappropriately selected: for a smaller aperture the outer annulus being differenced, to remove the background CMB, will be contaminated by cluster signal, while for the larger aperture regions with no cluster signal will be included in the averaged aperture temperature leading. In both cases we would anticipate an underestimated signal. For the CGC and WISE samples, focused a lower redshifts we find that the fixed and redshift-varying apertures give largely consistent results. For the CMASS+LOWZ sample that has a higher mean redshift, we find that, particularly for  separations between $\sim 100-140 Mpc$, the two aperture selection choices give results that vary by $\sim1\sigma$. The redshift scaled aperture also has larger statistical errors.

 \subsection{$\hat p_{kSZ}$ signal-to-noise estimates}
 \label{sec:analysis:SNR}
 
 For each dataset we calculate the signal-to-noise ratio (SNR) \cite{Hartlap:2006kj,DeBernardis:2016pdv}, 
 \begin{eqnarray}
 \left(\frac{S}{N}\right)^2 &=& \sum_{ij} \hat{p}_{kSZ}(r_{i})Cov^{-1}(r_{i},r_{j})\hat{p}_{kSZ}(r_{j}).
 \end{eqnarray} 
 The inverse of an unbiased estimator for some statistical variable $x$ is in general not an unbiased estimator for $x^{-1}$ so we account for the jackknife covariance bias by the standard correction factor given by
 \begin{equation} 
 Cov^{-1} = \frac{(N_{JK} - N_{bins} - 2)}{(N_{JK} - 1)} Cov_{JK}^{-1}.
 \end{equation}

For CGC$\times$SEVEM, Ade et al. \cite{Ade:2015lza}   consider SNR results (which they denote $\chi^2_{null}$) for a subset of three bins, 15, 38 and 81 Mpc/h, and report 0.3$\sigma$ and 0.4$\sigma$ significance for 5 and 8' apertures, respectively. In our analysis, we considered three similar bins, 22, 54, 115 Mpc (for $h=0.7$ these would denote 15, 38 and 83 Mpc$/h$) with errors from the rotation method, and found SNR = 1.6 and 2.5, which can be expressed as 0.45 and 0.72$\sigma$ confidence limits in the $N_{bin}=3$-dimensional parameter space. 

We find bin selection does lead to variations in the SNR. For the 5' case, for example: if we change the third bin from 115 to 108 Mpc we find the SNR shifts from 0.45 to 0.36$\sigma$; while adding an additional bin, 87 Mpc or 97 Mpc, leads to an SNR of 1.52$\sigma$ or 0.25$\sigma$, respectively. 

As expected, we find that the SNR is markedly lower for JK-estimated errors, consistent with the comparative sizes of the errors shown in Fig \ref{fig3}. For the 5' and 8' apertures, and the subset of three bins, at 22, 54, 115 Mpc, the SNR is 0.13 and 0.26$\sigma$, and a similar variation depending on  bin selected.

To compare  the SNR for each of the datasets in the paper, we avoid an arbitrary subset selection, and consider all bins in the range of comoving separations between 15 and 150 Mpc  and use the full covariance between each bin estimated with the JK method. We find that the highest SNR corresponds to the aperture size choice that most closely reflects an expected cluster size, as given in Table \ref{tab1}, but is not improved by using the redshift-dependent aperture for the cases with galaxy samples with extended redshifts. For the CGC data, we find SNR of $0.43\sigma$ and $1.0\sigma$ for the 5' and 8' apertures respectively. For CMASS+LOWZ, we find 0.05 and $0.37\sigma$ significance depending whether the redshift-dependent or 3' fixed aperture was assumed. For WISE, we similarly find an SNR of 0.02 and 0.04$\sigma$  for the redshift-dependent and fixed 5' aperture.

\begin{figure*}
	{
		\includegraphics[width=0.495\textwidth]{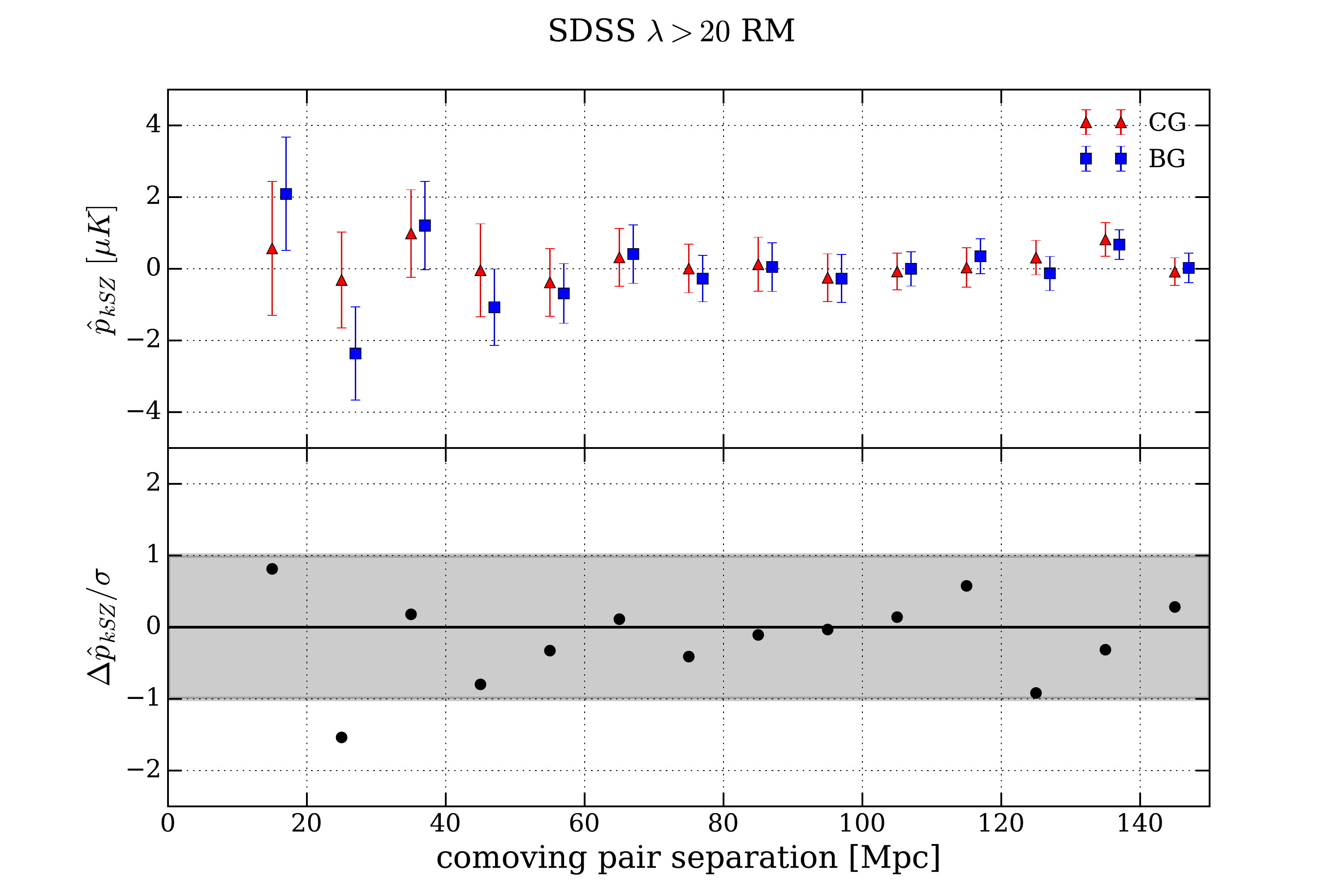}
		\includegraphics[width=0.495\textwidth]{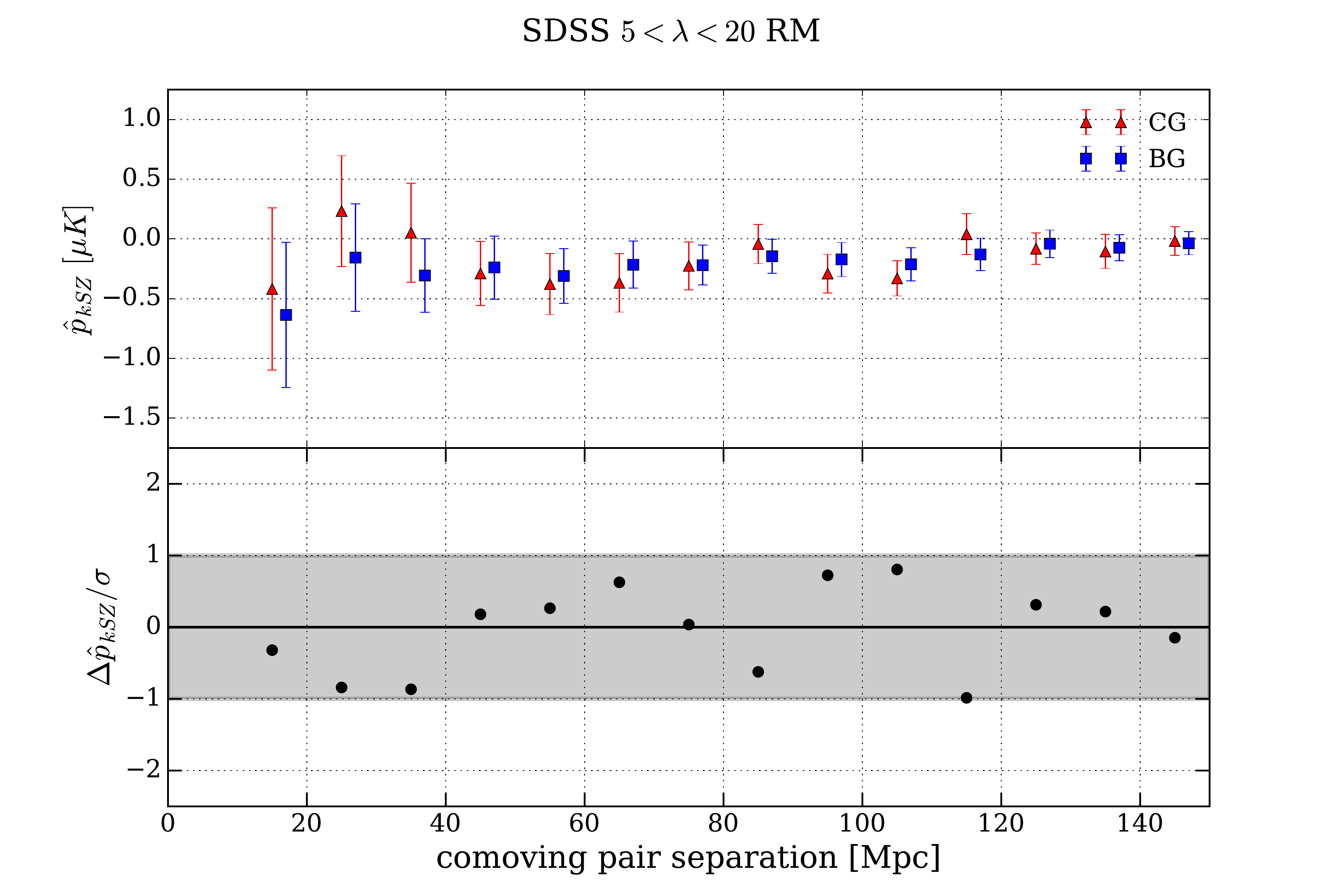}
	}
	\caption{$\hat{p}_{kSZ}$  for cross-correlations using  SDSS redMaPPer samples for clusters samples with [left panels] $\lambda>20$ and [right panels] $\lambda<20$, as summarized in Table \ref{tab1}. [Top panels] A comparison of $\hat{p}_{kSZ}$ derived using Central Galaxy catalogs (CG) [red triangle] and corresponding redMaPPer Brightest Galaxy (BG) [blue square] selections. [Lower panels] $\Delta \hat p_{kSZ}/\sigma$ is the difference in signal, $\hat{p}_{kSZ, BG}-\hat{p}_{kSZ, CG}$, scaled relative to the JK errors for the CG catalog, to measure the deviation of the signal between the brightest and true central galaxy, relative to the statistical error estimate.}
	\label{fig6}
\end{figure*}

\subsection{Impact of transverse miscentering}
\label{sec:analysis:centering}
In this section we discuss the impact of angular (transverse) offsets of the targeting galaxy from the cluster's center using methods described in \S\ref{sec:form:center}.

To create catalogs to further investigate the offset distributions, we take into account the five most likely central galaxy candidates based on the redMaPPer centering probability \cite{Hoshino} (considering those candidates with the highest $P_{cen,i}$, where $i =$ 1 through 5). 

In Figure~\ref{fig6}, we compare the results for the pairwise estimator found with the both a redMaPPer Brightest Galaxy (BG) catalog (the brightest LRG in the cluster) and a redMaPPer Central Galaxy (CG) catalog (the LRG with the highest centering probability) for moderate to high, $\lambda>20$, and low richness $5<\lambda<20$ cluster samples. In both cases, the results demonstrate that the uncertainty due to miscentering is comparable to  the JK statistical error estimates. We find deviations averaged over comoving separations $\sim$15 $-$ 155 Mpc to be $0.5\sigma$ for both the low and high richness SDSS samples, and a maximum deviation of $0.9\sigma$ at 35 Mpc and $1.5\sigma$ at 25 Mpc, for the low and high richness samples respectively.

As a complementary study, we use the photometric redMaPPer catalog directly, to study miscentering using the Johnston analytic model with the spectroscopic galaxy samples. We  inform the Johnston model parameters  using the RedMaPPer (RM) data. For our RM SDSS samples, we find $f_J$ is 70 $-$ 75\%, indicating that selecting brightest galaxy accurately identifies the cluster center 70 $-$ 75\% of the time. The value of $\sigma_J$ is estimated by computing $\doff$ and $P(\doff)$ based on the off-centering distribution given by

\begin{equation}
 P({\doff}_{ij}) = P_{cen,i}P_{mem,j} \Pi_k (1-P_{mem, k}),
\end{equation}

in which we normalize by the richness-dependent cutoff radius $R_c(\lambda)$ in equation (\ref{eq:Rc}), and fitting those to the $f_J$ given by the Johnston model above. This yields estimates of the $\sigma_J$ in units of Mpc$/h$: $0.28$ for the SDSS RM with $\lambda >20$, $0.35$ for the SDSS RM with $\lambda <20$, and $0.34$ for the LOWZ+CMASS DECaLs sample. 

Based on these results, we consider $\sigma_J=$ 0.3 Mpc$/h$, and 0.5 Mpc$/h$, $f_J=0.75$ and $h=0.6731$.

\begin{figure*}[!t]
	{
		\includegraphics[width=0.329\textwidth]{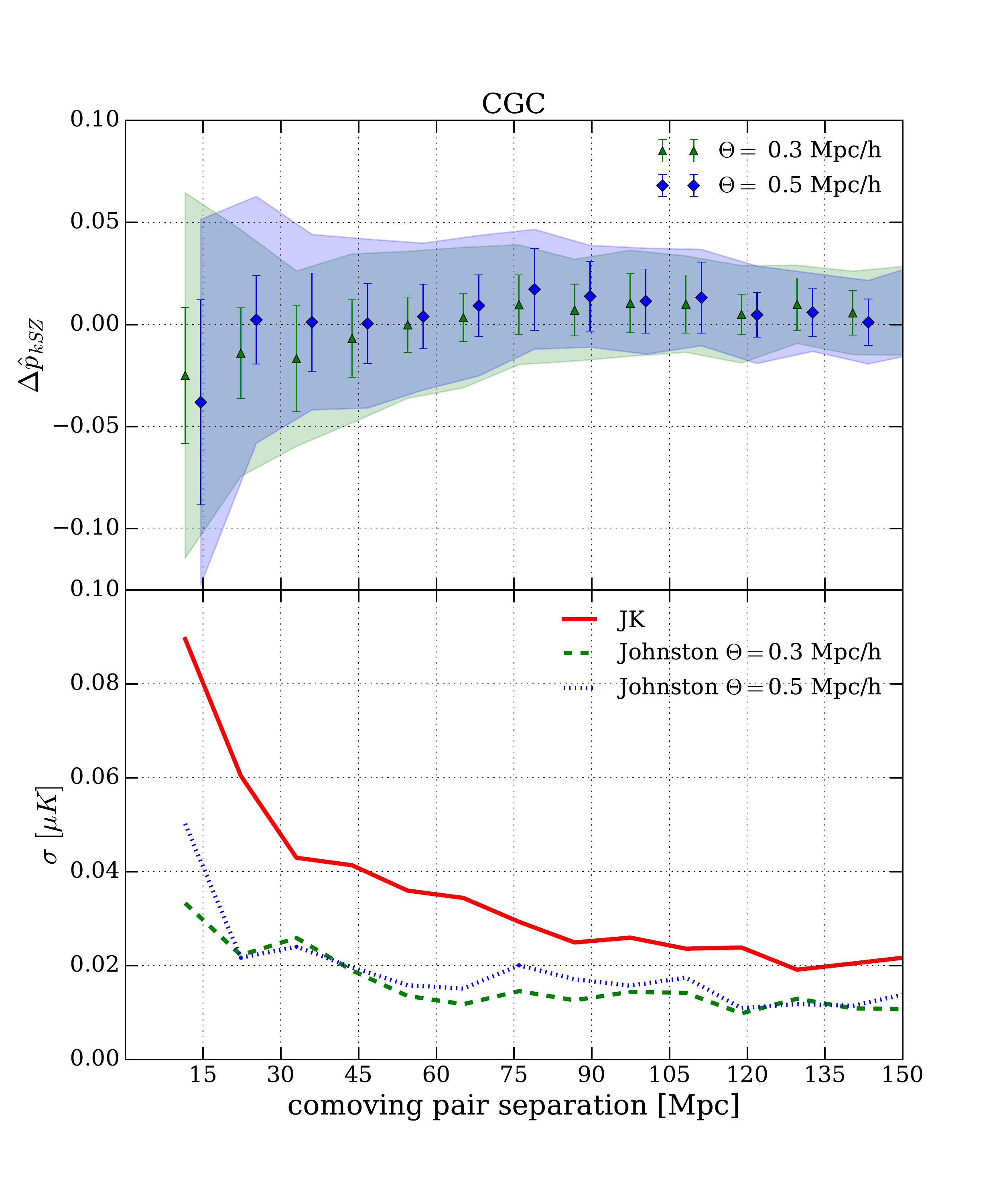}
		\includegraphics[width=0.329\textwidth]{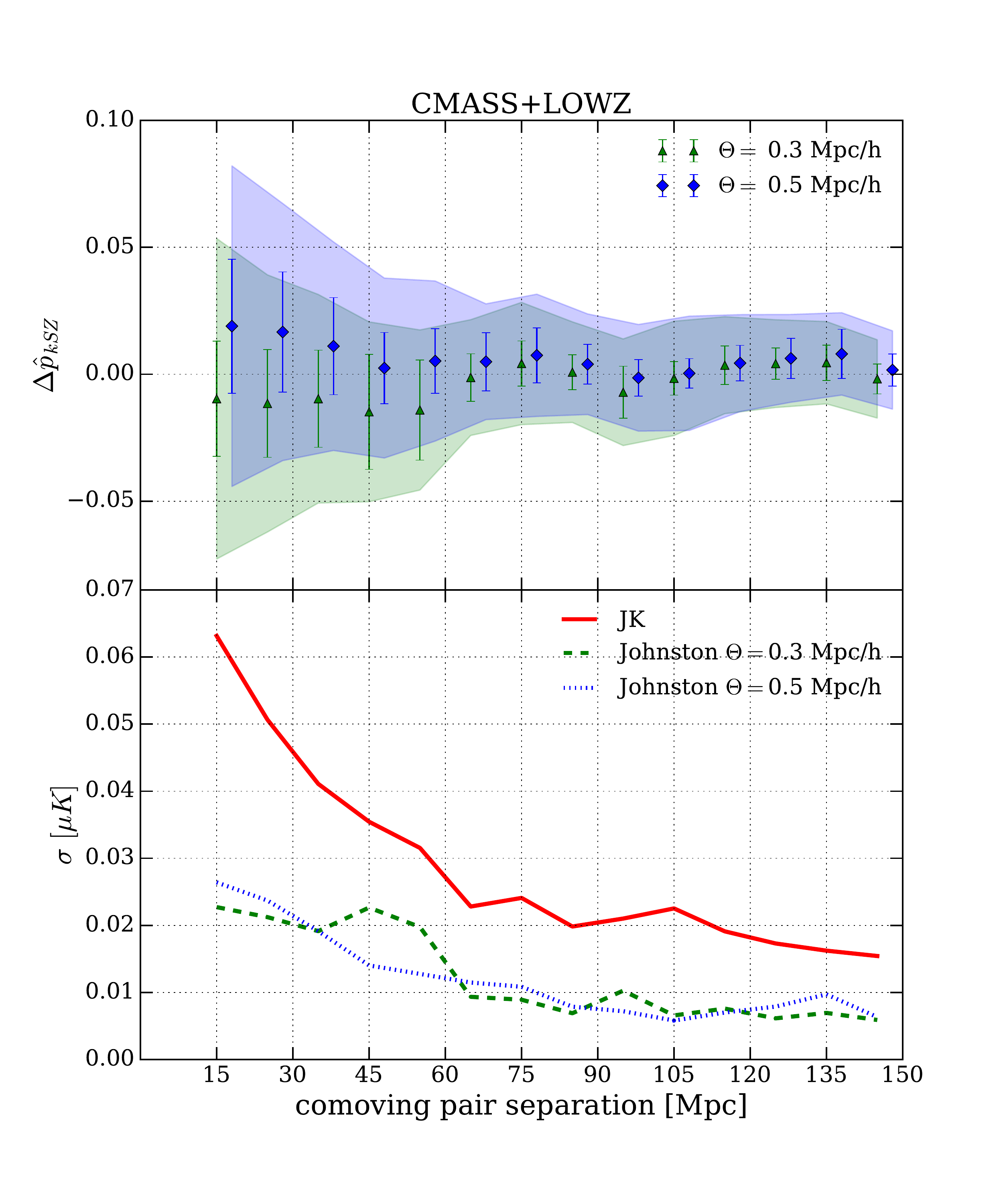}
		\includegraphics[width=0.329\textwidth]{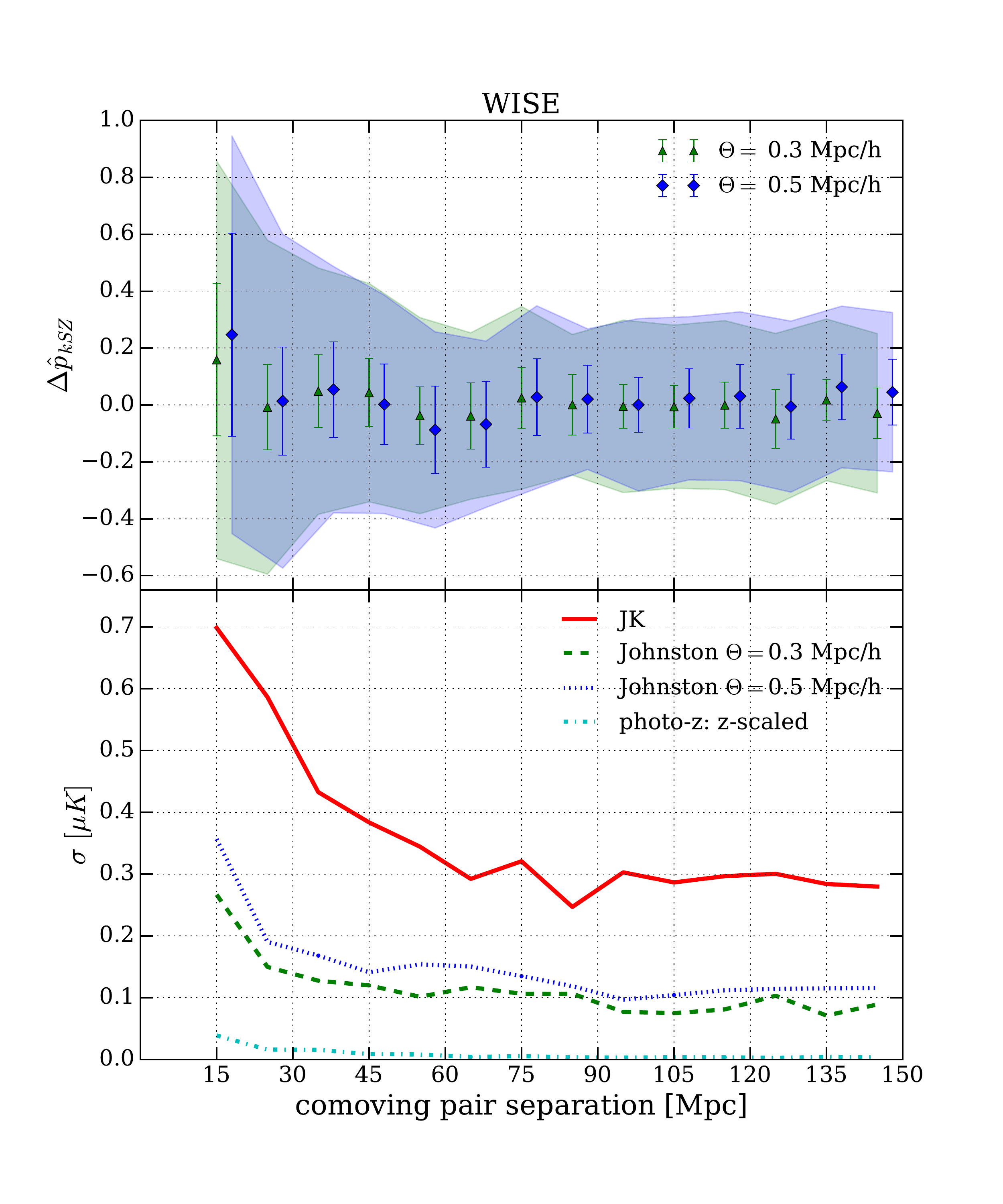}
	}
	\caption{A comparison of uncertainties arising from the Johnston miscentering model relative to JK errors for the [left] CGC, [center] CMASS+LOWZ, and [right] WISE samples, using aperture choices of 8 arcmin, 3 arcmin, and $z$-scaled	respectively. [Upper panel] $\Delta \hat p_{kSZ}$ is computed by $ \langle \hat p_{kSZ, offset_n} - \hat p_{kSZ} \rangle$, where $n$ is each of the 50 trials for $\sigma_J = 0.3$ [green triangle] and $0.5$ Mpc$/h$ [blue diamond]. The variance for each offset size is the r.m.s. difference between the individual 50 offset trials and the original signal. The shaded regions represent the JK variance of the original signal. [Lower panel] A comparison of the variance due jackknife resampling [red full], the Johnston offset model with an offset width 0.3 Mpc$/h$ [green dashed] and 0.5 Mpc$/h$ [blue dotted], for the original brightest-galaxy samples.}
	\label{fig7}
\end{figure*}

\begin{figure}
	{
		\includegraphics[width=0.5\textwidth]{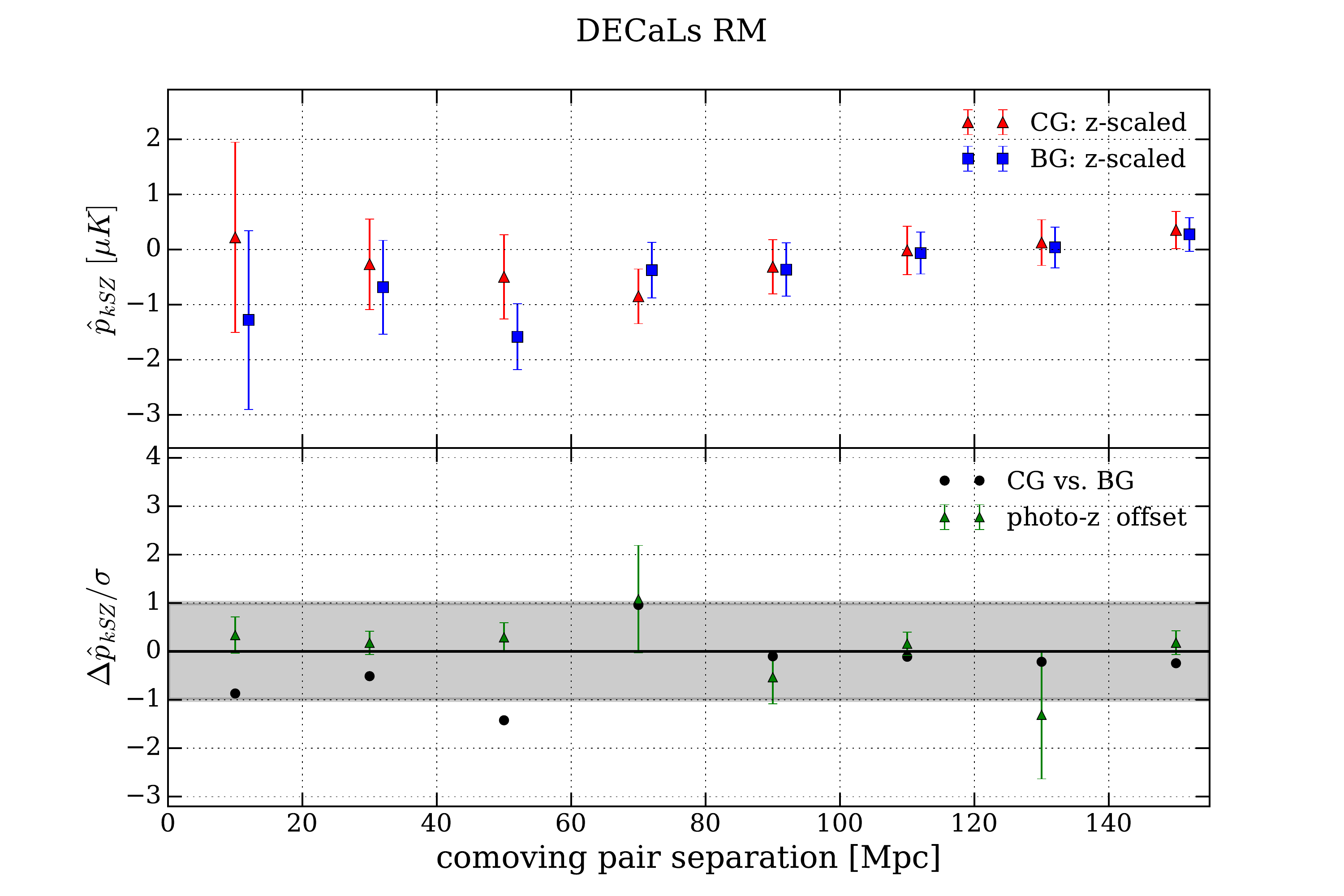}
	}
	\caption{A comparison of the kSZ signal from the brightest galaxy (BG) and central galaxy (CG) selections from DECaLs redMaPPer data using a $z$-dependent aperture. As in Figure~\ref{fig6}, [Top panel] a comparison of $\hat{p}_{kSZ}$ derived using Central Galaxy catalogs (CG) [red triangle] and corresponding redMaPPer Brightest Galaxy (BG) [blue square] selections. [Lower panel] $\Delta \hat p_{kSZ}/\sigma$ is the difference in signal, $\hat{p}_{kSZ, BG}-\hat{p}_{kSZ, CG}$, scaled relative to the JK errors for the CG catalog [black circle]. The lower panel also demonstrates the variance introduced by photometric redshift (`photo-z') errors on the kSZ signal. 50  trials are created with random offsets  to cluster locations, based a Gaussian  width corresponding to the photometric redshift error given for the CG sample. The average signal and variance of the 50 trials is shown as a ratio of the JK statistical errors [green triangle].}
	\label{fig8}
\end{figure}

We determine the relative importance of miscentering systematics to the statistical uncertainties by considering the covariances introduced in 50 offset trials, based on the same generating seed, in which a different $(1-f_J)=$ 25\% of the cluster locations shifted by the same random Gaussian displacement scaled by, standard deviation, $\sigma_J$.  For WISE,  only galaxies with $z > 0.03$ are  shifted, as the characteristics of the very low redshift sources are better known and miscentering is expected to be less of an issue, while concurrently the model would induce large angular offsets.

In Figure \ref{fig7}, we present the results of these offset trials for different LSS and CMB data pairings. The  variance in the pairwise signal between each of the 50 runs generated by the offsets are compared to the JK statistical errors from the jackknife analyses. The results show that, while miscentering does not introduce a systematic shift in the pairwise signal,  it does induce a significant enhancement to the kSZ statistical error budget. Averaged over the pairwise statistic at all separations ($\sim15 - 150$ Mpc), for CGC we find miscentering uncertainties are $0.7\sigma$ for $\sigma_J=0.3 Mpc/h$. For the CMASS+LOWZ, we find $0.6\sigma$ and for WISE, $0.4\sigma$. We find the maximum deviations occur for the CGC at $\sim$ 100 Mpc, of $0.7\sigma$, for WISE at 65 Mpc, of $0.5\sigma$, and for CMASS+LOWZ $0.8\sigma$ at 155 Mpc. Deviations at small separations where the kSZ effect is strongest are most important as a source of systematic uncertainty and in all cases the variation in $ \hat p_{kSZ}$ due to miscentering is significantly greater below separations of 60 Mpc. It is interesting to note that while both JK and miscentering errors decrease with increased sample size, progressing from WISE, to CGC and CMASS+LOWZ, the miscentering uncertainties concurrently become a larger fraction of the error budget, suggesting that they will not be ameliorated in future surveys simply as a result increased galaxy samples.
\subsection{Impact of photometric redshift errors}
\label{sec:analysis:pz}

 Photometric redshift errors can themselves be a source of contamination to the kSZ signal.  To analyze this prospect with the DECaLs RMCG sample 50 realizations were created in which galaxy radial positions had random offsets applied to the approximately half of the sample which have photometric instead of spectroscopic redshifts, sampled from a Gaussian of width corresponding to the  error given for the photometric redshifts. Figure \ref{fig8} gives a comparison of the variance in the realizations modeling the photometric redshift errors to the differences induced by shifting from the most likely central galaxy (CG) to the brightest galaxy (BG) catalog and the JK statistical errors. At comoving separations exceeding 175 Mpc, where the kSZ signal becomes approximately null, and the JK errors are very small, photometric redshift errors dominate.

At smaller separations, $<\sim$50 Mpc, photometric redshifts and miscentering both make significant contributions to the error budget, with differences between the BG and CG data being the larger. In the range of separations up to 50 Mpc, miscentering errors are comparable to the JK statistical errors, while photometric redshift errors constitute an average $0.28\sigma$, with a peak value of $0.34\sigma$ at 10 Mpc. 
\section{Conclusions}
\label{sec:conclusions}

The kSZ temperature deviation in galaxy clusters could provide a powerful probe of dark energy and modifications to gravity on a cosmic scale, distinct from galaxy lensing and clustering, that are principal science drivers for upcoming large scale structure surveys. Current approaches center on kSZ pairwise momentum estimates  between  clusters, obtained through cross-correlation of the CMB data with  galaxy samples that are used as cluster proxies to determine accurate pairwise separations. Improved resolution and frequency coverage in the next generation of sub-arcminute scale CMB measurements, twinned with  increased breadth and depth in  the next generation of LSS surveys, pave the way for significant improvements in kSZ signal extraction, but only if astrophysical and analysis systematics are understood and mitigated    at a  commensurate  level. 

In this paper, the impact of a number of modeling assumptions and  potential systematic uncertainties that can be introduced in the estimation of the pairwise kSZ correlation have been considered. Planck and WMAP CMB data are used in combination with galaxy samples from SDSS, WISE, and DECaLs surveys. 

In comparing covariance estimation techniques, the JK method was found to be more conservative than random rotations. A  sensitivity to aperture size selection was found for both methods, and  most pronounced for the galaxy samples distributed over  extended redshift ranges.  In contrast, a comparison of flat- vs. noise-weighting CMB maps was found to have little impact on the pairwise statistic from current Planck data.  
We also found negligible differences in  results  for the  SEVEM and LGMCA maps, for the same galaxy sample, suggesting that the aperture photometry method was  robust to the differences in residual foreground removal. 

The signal-to-noise ratio was evaluated for the CGC, CMASS+LOWZ and WISE samples. To provide context with other analyses in the literature, we considered the sensitivity to the selection of subsets of data and to aperture size. While none of the datasets leads to a new significant detection of the kSZ effect, we did find that the greatest SNR was when the fixed aperture size most closely reflect the expected angular size of a cluster for the sample.

The impact of miscentering was considered  using two complementary techniques: the Johnston analytical offset model and a redMaPPer-based comparison of signals assuming samples of the brightest (BG) versus the most likely central (CG) galaxy per cluster. The redMaPPer data was  used to inform the parameter choices used in Johnston model \cite{Johnston:2007uc}; for all samples, we found that $\sim$25\% of the predicted clusters had brightest galaxies that were offset with a Raleigh distribution with a peak $\sim$0.3 Mpc$/h$. In both the direct comparison of BG and CG redMaPPer catalogs and Johnston analytic model, miscentering leads to additional  uncertainties equivalent to a significant fraction of the JK error budget. Using redMaPPer, we find deviations  averaged over comoving separations $\sim15-150$ Mpc to be $0.5\sigma$ for both the low and high richness SDSS samples. Using Johnston offset modeling, with mean offset  $0.3$ Mpc$/h$,  we find $0.4\sigma$ for WISE, and $0.6\sigma$ for CMASS+LOWZ, and $0.7\sigma$ for CGC.

The DECaLs redMaPPer sample was used to compare photometric redshift and miscentering errors in tandem. 
Miscentering was found to be the dominant of the two  uncertainties at $<50$ Mpc, where the kSZ signal is largest, with deviations at the $\sim\sigma$ level. Photometric redshift errors were also not negligible however, as noted in \cite{Soergel:2016mce} and \cite{Keisler:2012eg}, with a mean deviation of $0.3\sigma$. 
 
This work provides quantitative evidence that uncertainties in cluster centering (in terms of both transverse and radial (redshift) location) can introduce significant uncertainties, comparable to current statistical errors in the kSZ pairwise signal. Order of magnitude improvements in instrumental precision and survey size anticipated with the next generation of CMB and LSS surveys will allow smaller separations, at which the kSZ signal is largest, to be accessible with photometric surveys. Miscentering uncertainties will need to mitigated, however, so as to not dominate the error budget. This work suggests a combination of spectroscopic redshift precision and multiple cluster galaxy populations from photometric surveys, such as will be obtained with Euclid and WFIRST, and in overlapping regions of the DESI and LSST surveys, may be optimal to appropriately constrain both transverse and radial cluster positioning. Similarly this suggests implications for miscentering in kSZ analyses with simulations based on upcoming surveys due to the fact that even the true brightest cluster galaxy does not always trace the location of the projected center of electron distribution.
 
Looking forward, a variety of techniques may be employed to better extract the kSZ signal, such as the application of  matched-filter estimators \cite{2014MNRAS.443.2311L} and Gaussian constrained realizations \cite{2016arXiv160401382A} that will improve kSZ decrement estimation and also reduce reliance on cluster centering. Improved precision in cluster centering may also be achieved through using additional data, such as weak lensing information \cite{Hikage:2012mnr}.

Going beyond this, multi-frequency CMB temperature and polarization data will provide opportunities to extend from precise  pairwise momenta measurements to both pairwise, and individual, cluster velocities to fully realize the potential of the kSZ effect for cosmology. This will require concurrent measurements of cluster optical depth with an additional set of astrophysical and analysis assumptions that can contribute to the error budget \cite{Battaglia:2016xbi,Mittal:2017hwf}. While we have considered pairwise correlations for comoving separations between 15 and 150 Mpc, to fully understand the impact of analysis assumptions on signal extraction at scales where the signal is largest, clearly other considerations such as velocity biasing \cite{Baldauf:2014fza} and non-linear clustering will also need to be understood and characterized in order to convert the extracted signal to a peculiar velocity and cosmological model. Alternative approaches, such as using three-point statistics of all large scale structure \cite{Hill:2016dta, 2016PhRvD..94l3526F}, also open up kSZ science that may not rely on precise cluster identification. We leave detailed consideration of these next steps to future work. 

\section*{Acknowledgments}
We thank Eduardo Rozo and Eli Rykoff for providing the redMaPPer catalogs used in this analysis, and Wenting Wang and Carlos Hern\'{a}ndez-Monteagudo for sharing a CGC-derived catalog developed for the Planck kSZ analysis in \cite{Ade:2015lza}. We are grateful to Nicholas Battaglia, Francesco De Bernardis,  Hanako Hoshino, Michael Niemack and Eve Vavagiakis for helpful conversations in the course of this work. The work of VC and RB is supported by NASA ATP grant NNX14AH53G, NASA ROSES grant 12-EUCLID12- 0004 and DoE grant DE-SC0011838. 
\clearpage
\bibliographystyle{apsrev}

\end{document}